\documentclass[a4paper,aps,prd,twocolumn,tightenlines,preprintnumbers,nofootinbib,showkeys,superscriptaddress]{revtex4-1}
\pdfoutput=1

 \usepackage{XCharter}
 \usepackage[T1]{fontenc}
 \usepackage{mathptmx}

\usepackage{fullpage}
\usepackage{amsfonts}
\usepackage{amsmath}
\usepackage{slashed}
\usepackage{amssymb}
\usepackage{graphicx}
\usepackage{epic}
\usepackage{eepic}
\usepackage{epsfig}
\usepackage{latexsym}
\usepackage[dvipsnames]{xcolor}
\usepackage[export]{adjustbox}
\usepackage{float}
\usepackage{multirow}
\usepackage{hyperref}
\usepackage{enumitem}
\hypersetup{colorlinks=true,citecolor=red,linkcolor=NavyBlue,urlcolor=NavyBlue}
\usepackage[caption=false]{subfig}
 
\usepackage{natbib}
\usepackage{relsize}
\usepackage[left=1.75cm,right=1.75cm,top=2.1cm,bottom=2.5cm]{geometry}
\usepackage{shorthand}

\begin{document}
\relscale{1.05}

\title{Improving third-generation leptoquark searches with combined signals and boosted top quarks}

\author{Arvind Bhaskar}
\email{arvind.bhaskar@research.iiit.ac.in}
\affiliation{Center for Computational Natural Sciences and Bioinformatics, International Institute of Information Technology, Hyderabad 500 032, India}

\author{Tanumoy Mandal}
\email{tanumoy@iisertvm.ac.in}
\affiliation{Indian Institute of Science Education and Research Thiruvananthapuram, Vithura, Kerala, 695 551, India}

\author{Subhadip Mitra}
\email{subhadip.mitra@iiit.ac.in}
\affiliation{Center for Computational Natural Sciences and Bioinformatics, International Institute of Information Technology, Hyderabad 500 032, India}

\author{Mohit Sharma}
\email{mohit.sharma@research.iiit.ac.in }
\affiliation{Center for Computational Natural Sciences and Bioinformatics, International Institute of Information Technology, Hyderabad 500 032, India}

\date{\today}

\begin{abstract}
\noindent
Search strategies for the third-generation leptoquarks (LQs) are distinct from other LQ searches, especially when they decay to a top quark and a $\tau$ lepton. We investigate the cases of all TeV-scale scalar and vector LQs that decay to either a top-tau pair (charge-$1/3$ and charge-$5/3$ LQs) or a top-neutrino pair (charge-$2/3$ LQs). One can then use the boosted top (which can be tagged efficiently using jet-substructure techniques) and high-$p_{\rm T}$ $\tau$ leptons to search for these LQs. We consider two search channels with either one or two taus along with at least one hadronically decaying boosted top quark. We estimate the high luminosity LHC (HL-LHC) search prospects of these LQs by considering both symmetric and asymmetric pair and single production processes. Our selection criteria are optimised to retain events from both pair and single production processes. The combined signal has better prospects than the traditional searches. We include new three-body single production processes to enhance the single production contributions to the combined signal. We identify the interference effect that appears in the dominant single production channel of charge-$1/3$ scalar LQ ($S^{1/3}$). This interference is constructive if $S^{1/3}$ is a weak-triplet and destructive for a singlet one. As a result, their LHC prospects differ appreciably.
\end{abstract}

\maketitle


\section{Introduction}
\label{sec:intro}
\noindent
Recently leptoquarks (LQs) are receiving a lot of attention in the literature. Since they connect the quark and the lepton sectors of the Standard Model (SM), their phenomenology is interesting in general. 
However, the recent attention is mainly due to their ability to explain the long-standing anomalies observed in $B$-meson decays and the muon anomalous magnetic moment measurements (see, e.g.,~\cite{Sahoo:2016pet,Dey:2017ede,Bandyopadhyay:2018syt,Dorsner:2018ynv,Crivellin:2018yvo,Mandal:2018kau,Biswas:2018snp,
Biswas:2018iak,Alves:2018krf,Aydemir:2019ynb,Chandak:2019iwj,
Mandal:2019gff,Hou:2019wiu,Allanach:2019zfr,
Crivellin:2019dwb,Padhan:2019dcp,Bhaskar:2020kdr,Bandyopadhyay:2020klr,Bhaskar:2020gkk,Dev:2020qet,Buonocore:2020erb,Bandyopadhyay:2020wfv,Babu:2020hun,Crivellin:2020mjs,Greljo:2020tgv,Haisch:2020xjd,Bandyopadhyay:2020jez,Crivellin:2021egp,Dorsner:2021chv,Angelescu:2021lln,Hiller:2021pul,Asadi:2021gah,Bhaskar:2021pml,Bhaskar:2021owe} 
for a few recent studies). 
Generally, the models explaining the anomalies contain LQs that couple with the third-generation quarks and leptons. This gives additional motivation to search for such LQs (commonly known as the third-generation LQs in the literature) at the Large Hadron Collider (LHC).

Various strategies have already been applied in the searches for  third-generation LQs at the LHC~\cite{Gripaios:2010hv,Sirunyan:2018ruf,Sirunyan:2018vhk,Aaboud:2019bye,Aad:2020jmj,Sirunyan:2020zbk,ATLAS:2021aui,ATLAS:2021oiz}. In this paper, we consider a new search strategy for the third-generation scalar and vector LQs (sLQs and vLQs) that couple to a top quark and produce a boosted-top signature. We consider two signatures to estimate their discovery/exclusion prospects at the high-luminosity LHC (HL-LHC)---either one or two $\tau$-leptons in association with at least one hadronically-decaying boosted top quark. This paper, thus, complements our previous prospect studies on LQs that dominantly decay to a top quark and a light charged lepton (either $e$ or $\mu$), producing a top-lepton resonance~\cite{Chandak:2019iwj,Bhaskar:2020gkk}. The $\tau$-leptons are different from light leptons---they are unstable, and their decays always involve some missing energy making the $\tau$-tagging efficiency lower than that for the light leptons. Therefore, analysing the signatures involving $\tau$-leptons needs additional care.

 The two-$\tau$ signature is for charge-$1/3$ and charge-$5/3$ LQs, whereas the one-$\tau$ signature is for LQs with charge $2/3$ and $1/3$. We look for at least one hadronically-decaying boosted top-jet that can be tagged with the jet-substructure technique in both cases. Using this technique, one can tag a top-jet with good significance and enhance the LHC reach.
In a previous paper~\cite{Mandal:2015vfa}, we argued that it is possible to improve the LHC discovery prospect of LQs (or any other new coloured particles, for that matter) or exclusion limits if we systematically combine pair and single production channels. In this case, the two signatures capture combinations of pair and single production processes in the signal. Later, in Ref.~\cite{Aydemir:2019ynb}, we highlighted the importance of considering asymmetric channels for LQ searches. Here, we consider both symmetric and asymmetric channels and optimise the selection criteria to retain a substantial fraction of pair and single production events. We note, recently, the CMS Collaboration has also used a similar asymmetric channel ($tb\tau\nu$ signature) and combined the pair and single production channels in their search for the third-generation LQs~\cite{Sirunyan:2020zbk}.

Generally,  the simplest two-body single production processes are considered for the LQ searches in single production channels. However, there are three-body single production processes of LQs that are also important and have comparable or larger cross sections than the two-body single productions~\cite{Mandal:2015vfa}. In this paper,
we consider the three-body single productions and show that the combined signal can lead to better exclusion or discovery prospects.

Since the single-production cross sections depend on the LQ couplings, our results depend on LQ models. We have found an important point; for the same strength of the LQ couplings and decays, the HL-LHC discovery reach for the charge-$1/3$ component of the triplet scalar $S_3$ is higher than that of the singlet $S_1$, another charge-$1/3$ sLQ. This happens because of constructive interference among some subprocesses in the $pp\to S_3^{1/3}\tau j$ single production process. For $S_1$, this interference is destructive in nature. Obviously, a similar difference also appears in the exclusion limits of these two sLQs. Hence, one has to be mindful about the type of LQ and look beyond branching ratios while reading the exclusion limits on charge-$1/3$ sLQs. 
This interference effect is observed only in some channels and in case of charge-$1/3$ LQ models; other LQ species we have considered in this paper do not show such an effect.

Before we proceed further, we note that since this paper is a follow up to Refs.~\cite{Chandak:2019iwj,Bhaskar:2020gkk}, we  frequently refer to these papers and omit some common details while ensuring our current presentation is self-contained. The rest of the paper is organized as follows. In Sec.~\ref{sec:model} we describe the LQ models and discuss simplified models suitable for experimental analysis. In Sec.~\ref{sec:pheno}, we discuss
the LHC phenomenology and illustrate our search strategy, and then we present our estimations in Sec.~\ref{sec:dispot}. Finally, we summarise
and conclude in Sec.~\ref{sec:End}.


\begin{table*}[t]
\centering{\linespread{1.5}
\begin{tabular*}{\textwidth}{l @{\extracolsep{\fill}}|lc @{\extracolsep{\fill}}ccc  @{\extracolsep{\fill}}cc @{\extracolsep{\fill}}cc}\hline
\multicolumn{2}{c}{}&&\multicolumn{3}{c}{Simplified models [Eqs.~\eqref{eq:simpleSlag1} -- \eqref{eq:simplevlag5}]}&\multicolumn{2}{c}{LQ models [Eqs.~\eqref{eq:LagU1t} -- \eqref{eq:LagU3p}]}&&\\\cline{4-6}\cline{7-8}\multicolumn{2}{c}{
\begin{tabular}[c]{l}Benchmark \\ scenario\end{tabular}} & \begin{tabular}[c]{c}Possible \\ charge(s)\end{tabular} & \begin{tabular}[c]{c}Type of  \\ LQ\end{tabular} & \begin{tabular}[c]{c}Nonzero \\ couplings\\ equal to $\lm/\Lm$\end{tabular} & \begin{tabular}[c]{c}Charged\\lepton \\ chirality \\fraction\end{tabular} & \begin{tabular}[c]{c}Type of  \\ LQ\end{tabular} &\begin{tabular}[c]{c}Nonzero \\ coupling\\ equal to $\lm/\Lm$\end{tabular} & \begin{tabular}[c]{c}Decay \\ mode(s)\end{tabular}& \begin{tabular}[c]{c}Branching \\ ratios(s)\\$\{\bt,1-\bt\}$\end{tabular}\\
\hline\hline\multirow{8}{*}{
\rotatebox[origin=lB]{90}{Scalar~~~~~~}
}
&\multirow{2}{*}{LC}      		&$2/3$		& $\phi_2$  & $\hat\lm_\nu$		&---		& $\left\{\left(S_3^{-2/3}\right)^\dag, R_2^{2/3}\right\} $	 		&$\left\{\sqrt2\left(y^{LL}_{3\ 33}\right)^*,y^{RL}_{2\ 33}\right\}$		&  $t\nu$ &  \multirow{2}{*}{$\{100\%,0\}$}\\
&       &$5/3$		& $\phi_5$  & $\tilde\lm_\tau$		& $\eta_L=1$ 		& $R_2^{5/3}$	 		&$-y_{2\ 33}^{RL}$		&  $t\tau$ &   \\
\cline{2-10}
&LCSS   	&\multirow{2}{*}{$1/3$}			& \multirow{2}{*}{$\phi_1$}    			&	$\lm_\tau=\lm_\n$         	& \multirow{2}{*}{$\eta_L=1$}& $S_3^{1/3}$     					&$-y^{LL}_{3\ 33}$		& \multirow{2}{*}{\{$t\tau$, $b\n$\}} 	& \multirow{2}{*}{\{$50\%$, $50\%$\}}  \\
&LCOS   	&			&      			& $\lm_\tau=-\lm_\n$ 			& & $S_1$    					&$y_{1\ 33}^{LL}$ 		&   &         \\\cline{2-10}
&\multirow{2}{*}{RC}      		&$1/3$		& $\phi_1$  & $\lm_\tau$		& $\eta_R=1$ 		& $S_1$	 		&$y_{1\ 33}^{RR}$		&  $t\tau$ &  \multirow{2}{*}{$\{100\%,0\}$} \\
&      		&$5/3$		& $\phi_5$  & $\tilde\lm_\tau$		& $\eta_R=1$ 		& $R_2^{5/3}$ 	 		&$y^{LR}_{2\ 33}$		&  $t\tau$ &  \\\cline{2-10}
&RLCSS*  	&\multirow{2}{*}{$2/3$}	   & \multirow{2}{*}{$\phi_2$}	& $\hat\lm_{\tau}=\hat\lm_{\nu}$                & \multirow{2}{*}{$\eta_R=1$}	& $\displaystyle R_2^{2/3}$     					&$y_{2\ 33}^{RL}=\left(y_{2\ 33}^{LR}\right)^*$  & \multirow{2}{*}{\{$t\nu$, $b\tau$\}} 	& \multirow{2}{*}{\{$50\%$, $50\%$\}}  \\
&RLCOS*  	&	   & 	& $\hat\lm_{\tau}=-\hat\lm_{\nu}$                & 	& $\displaystyle R_2^{2/3}$     					&$y_{2\ 33}^{RL}=-\left(y_{2\ 33}^{LR}\right)^*$  &&\\
\hline
\multirow{9}{*}{
\rotatebox[origin=lB]{90}{Vector~~~~~~}
}&\multirow{3}{*}{LC}      & $1/3$   &   $\chi_1$  &   $\Lm_{\tau}$        & $\eta_L=1$      &  $\displaystyle\tilde V_2^{1/3}$        & $\tilde x_{2\ 33}^{RL}$           & $t\tau$ & \multirow{3}{*}{$\{100\%,0\}$}\\
        && $2/3$  &   $\chi_2$  &   $\hat\Lm_\n$      &---                       & $\displaystyle\lt(\tilde V_2^{-2/3}\rt)^\dag$       & $\displaystyle\lt(\tilde x_{2\ 33}^{RL}\rt)^*$           & $t\n$   &\\
        && $5/3$   &   $\chi_5$  &   $\tilde\Lm_\tau$  & $\eta_L=1$   &     $\displaystyle U_3^{5/3}$        & $\sqrt{2}\ x_{3\ 33}^{LL}$ & $t\tau$ & \\
\cline{2-10}
&LCSS*  	&\multirow{2}{*}{$2/3$}	   & \multirow{2}{*}{$\chi_2$}	& $\hat\Lm_{\tau}=\hat\Lambda_{\nu}$        & \multirow{2}{*}{$\eta_{L}=1$}&	 $\displaystyle U_1$     					&$x_{1\ 33}^{LL}$  & \multirow{2}{*}{\{$t\n$, $b\tau$\}}	& \multirow{2}{*}{\{$50\%$, $50\%$\}}  \\
&LCOS 	&	    &	& $\hat\Lm_{\tau}=-\hat\Lambda_{\nu}$        && $\displaystyle U_3^{2/3}$     					&$-x_{3\ 33}^{LL}$  &  	&  \\
\cline{2-10}
&\multirow{2}{*}{RC}        & $1/3$   &   $\chi_1$  &   $\Lm_\tau$        & \multirow{2}{*}{$\eta_R=1$}       & $\displaystyle V_2^{1/3}$               & $x_{2\ 33}^{LR}$                  & \multirow{2}{*}{$t\tau$}  & \multirow{2}{*}{$\{100\%,0\}$}\\
        && $5/3$   &   $\chi_5$  &   $\tilde\Lm_\tau$       &    & $\displaystyle\tilde U_1$              & $\tilde x_{1\ 33}^{RR}$           & &\\
\cline{2-10}
&RLCSS*  	&\multirow{2}{*}{$1/3$}	   & \multirow{2}{*}{$\chi_1$}	& $\Lambda_{\tau}=\Lambda_{\nu}$                & \multirow{2}{*}{$\eta_{R}=1$}	& $\displaystyle V_2^{1/3}$     					&$x_{2\ 33}^{LR}=x_{2\ 33}^{RL}$  & \multirow{2}{*}{\{$t\tau$, $b\n$\}} 	& \multirow{2}{*}{\{$50\%$, $50\%$\}}  \\
&RLCOS*  	&	   & 	& $\Lambda_{\tau}=-\Lambda_{\nu}$                & 	& $\displaystyle V_2^{1/3}$     					&$x_{2\ 33}^{LR}=-x_{2\ 33}^{RL}$  &&\\
 \hline
\end{tabular*}}
\caption{Summary of the benchmark scenarios showing the map between the known LQ models [Eqs.~\eqref{eq:LagU1t} -- \eqref{eq:LagU3p}] and the simple models [Eqs.~\eqref{eq:simpleSlag1} -- \eqref{eq:simplevlag5}]. The branching ratio $\bt$ for a $\phi/\chi$ to decay to a top quark is fixed in all scenarios, except for $U_1$ in the vector LCSS scenario $(\bt\leq 50\%)$ and $R_2^{2/3}$ ($V_2^{1/3}$) in the scalar (vector) RLCSS/OS scenarios where $0\leq\bt<100\%$ [for $\bt=100\%$, the scalar (vector) RLCSS/OS scenarios become the same as the scalar LC (vector RC) scenario]. An asterisk marks exceptional scenarios. Here, $\lm/\Lm$ is a generic free-coupling parameter. We have chosen only this one coupling to control all of the nonzero new couplings in every benchmark for simplicity. This essentially means also choosing $\bt$ to be $50\%$ in the exceptional scenarios.}\label{tab:benchmark}
\end{table*}

\section{Scalar and vector leptoquark models}\label{sec:model}

\noindent
We are interested in
the LHC phenomenology of LQs that couple simultaneously to a top quark and a third-generation lepton.
The scalar and vector LQs that can have the desired couplings are {$S_1$, $R_2$ and $S_3$} and {$U_1$, $\bar{U}_1$, $V_2$, $\tilde{V}_2$ and $U_3$}, respectively (different symbols are used in the original papers~\cite{Blumlein:1994qd,Blumlein:1996qp} but the current symbols are taken from Ref.~\cite{Dorsner:2016wpm}). To avoid proton-decay constraints, we ignore the diquark operators. 
For simplicity, we also assume both the Pontecorvo-Maki-Nakagawa-Sakata neutrino-mixing matrix and the Cabibbo-Kobayashi-Maskawa quark-mixing matrix to be unity. This is justified since  the LHC is blind to the flavor of the neutrinos and the small off-diagonal terms in the CKM matrix would not play a significant role in our analysis.

\subsection{Scalar leptoquark models}

\label{sec:scalarmodels}
\noindent
$\blacksquare\quad$\underline{$S_1=(\overline{\mathbf{3}},\mathbf{1},1/3)$:}
The interaction Lagrangian of the sLQ $S_1$ can be written as follows:
\begin{align}
\label{eq:LagS1}
\mathcal{L} \supset &~ y^{LL}_{1\,ij}\bar{Q}_{L}^{C\,i} S_{1} i\sigma^2 L_{L}^{j}+y^{RR}_{1\,ij}\bar{u}_{R}^{C\,i} S_{1} \ell_{R}^{j}+\textrm{H.c.},
\end{align}
where $u_{R}$ and $\ell_{R}$ are a SM right-handed up-type quark and a charged lepton, respectively and  $Q_L$ and $L_L$ are the SM left-handed quark and lepton doublets, respectively. The superscript $C$  denotes  
charge conjugation and $\sigma^2$ is the second Pauli matrix. The generation indices are denoted 
by $i,j=\{1,\ 2,\ 3\}$. The color indices are suppressed. Since the LHC cannot distinguish between the flavours of the neutrinos, we write them as $\nu$.
The terms relevant for our analysis are given as
\begin{align}
\mathcal{L} \supset &\ y^{LL}_{1\ 33} \left(-\bar{b}_{L}^C \nu_{L}+\bar{t}_{L}^{C}\tau_{L} \right)S_{1}
+y^{RR}_{1\,33}\ \bar{t}_{R}^{C} \tau_{R} S_{1}+\textrm{H.c.}\label{eq:LagS1us}
\end{align}

\noindent
$\blacksquare\quad$\underline{$S_3= (\overline{\mathbf{3}},\mathbf{3},1/3)$:}
There is only one type of renormalisable operator one can write for $S_3$ that is invariant under the SM gauge group
\begin{align}
\label{eq:LagS3}
\mathcal{L} \supset &~y^{LL}_{3\,ij}\bar{Q}_{L}^{C\,i,a} \epsilon^{ab} (\tau^k S^k_{3})^{bc} L_{L}^{j,c} + \textrm{H.c.}
\end{align}
Here, the $\mathrm{SU}(2)$ indices are denoted 
by $a,b,c=\{1,\ 2\}$.  Expanding and keeping only the relevant interaction terms, we get,
\begin{align}
\mathcal{L} \supset  -y^{LL}_{3\ 33}&\left[\left(\bar{b}_{L}^{C} \nu_{L}+\bar{t}_{L}^{C} \tau_{L}\right) S^{1/3}_3+\sqrt{2}\left( \bar{b}_{L}^{C}  \tau_{L}S^{4/3}_{3}\right.\nn\right.\\
&-
\left.\left.\bar{t}_{L}^{C} \nu_{L}S^{-2/3}_3\right)\right]+\textrm{H.c.}\label{eq:LagS3us}
\end{align}

\noindent
$\blacksquare\quad$\underline{$R_2=(\mathbf{3},\mathbf{2},7/6)$:} Similarly, for $R_2$ we have the following terms,
\begin{eqnarray}\label{eq:LagR2}
\mathcal{L} &\supset&-y^{RL}_{2\,ij}\bar{u}_{R}^{i} R_{2}^{a}\epsilon^{ab}L_{L}^{j,b}+y^{LR}_{2\,ji}\bar{e}_{R}^{j} R_{2}^{a\,*}Q_{L}^{i,a} +\textrm{h.c.},\ \nn
\end{eqnarray}
which, after expansion and considering the relevant terms, can be written as
\begin{eqnarray}
\mathcal{L} &\supset&  - y^{RL}_{2\ 33}\ \bar{t}_R \tau_L R_2^{5/3}+  y^{RL}_{2\ 33}\ \bar{t}_R 
  \nu_L R_2^{2/3}\nn\\
 &&+y^{LR}_{2\ 33}\ \bar{\tau}_R 
  t_L R_2^{5/3\,*} +y^{LR}_{2\ 33}\ \bar{\tau}_R b_L R_2^{2/3\,*} + \textrm{H.c.}\label{eq:main_R_2_a}
\end{eqnarray}

\subsection{Vector leptoquark models}
\label{sec:vectormodels}
\noindent
$\blacksquare\quad$\underline{$\tilde{U}_{1}$ = (${\mathbf{3}}$,$\mathbf{1}$,$5/3$):}\quad
The electric charge of $\tilde{U}_{1}$ is $5/3$. Hence, it couples exclusively with the right-handed leptons:
\be \label{eq:LagU1t}
\mc L \supset \tilde{x}_{1~ij}^{RR}\bar{u}_{R}^{i}\gamma^{\mu}\tilde{U}_{1,\m} \ell_{R}^{j} + \textrm{H.c.},
\ee
For our purpose, we have
\begin{equation} \label{eq:LagU1tp}
\mathcal{L} \supset \tilde{x}_{1~33}^{RR}~\bar{t}_{R} \lt(\gm\cdot\tilde{U}_{1}\rt) \tau_{R} +  \textrm{H.c.}
\end{equation}

\noindent
$\blacksquare\quad$\underline{$U_{1}$ = ($\mathbf{3}$,$\mathbf{1}$,$2/3$):}\quad
The necessary interaction terms for the charge-$2/3$ $U_1$ can be written as
\ba 
\label{eq:LagU1}
\mathcal{L} \supset x_{1\ ij}^{LL}~\bar{Q}_{L}^{i}\gamma^{\mu}U_{1,\mu}L_{L}^{j}  + x_{1\ ij}^{RR}~\bar{d}_{R}^{i}\gamma^{\mu}U_{1,\mu}\ell_{R}^{j}+ \textrm{H.c.}~~
\ea
The $i,j=3$ terms can be written explicitly as
\begin{align}
\label{eq:LagU1p}
\mathcal{L} \supset& x_{1\ 33}^{LL}\left\{\bar{t}_{L}\lt(\gamma\cdot U_{1}\rt)\n_{L} + \bar{b}_{L}\lt(\gamma\cdot U_{1}\rt)\tau_L\rt\} \nn\\
&+x_{1\ 33}^{RR}~\bar{b}_{R}\lt(\gamma\cdot U_{1}\rt)\tau_R+\textrm{H.c.}
\end{align}

\noindent
$\blacksquare\quad$\underline{$V_{2}$ = ($\bar{\mathbf{3}}$,$\mathbf{2}$,$5/6$):}\quad
For $V_{2}$, the Lagrangian is as follows:
\begin{align}
\label{eq:LagV2}
\mathcal{L} \supset& x_{2\ ij}^{RL}~\bar{d}_{R}^{Ci}\gamma^{\mu} V_{2,\mu}^{a} \epsilon^{ab} L_{L}^{jb}\nn\\
&+ x_{2\ ij}^{LR}~\bar{Q}_{L}^{Ci,a}\gamma^{\mu}\epsilon^{ab} V_{2,\mu}^{b} \ell_{R}^{j}+
 \textrm{H.c.}
\end{align}
Expanding the Lagrangian we get the terms relevant for our analysis as,
\begin{align}
\mathcal{L} \supset & 
-x_{2\ 33}^{RL}\bar{b}_{R}^{C}\left\{\lt(\gamma\cdot V_{2}^{1/3}\rt)\n_L
- \lt(\gamma\cdot V_{2}^{4/3}\rt)\tau_{L}\rt\}\nn\\
&+x_{2\ 33}^{LR}\left\{\bar{t}_{L}^{C}\lt(\gamma\cdot V_{2}^{1/3}\rt)
- \bar{b}_{L}^{C}\lt(\gamma\cdot V_{2}^{4/3}\rt)\rt\}\tau_{R} + \textrm{H.c.} \label{eq:LagV2p}
\end{align}

\noindent
$\blacksquare\quad$\underline{$\tilde V_{2}$ = ($\bar{\mathbf{3}}$,$\mathbf{2}$,$-1/6$):}\quad
For $\tilde V_{2}$, the Lagrangian is given as,
\begin{align}
\label{eq:LagV2t}
\mathcal{L} \supset&\ \tilde{x}^{RL}_{2\ ij}\bar{u}_{R}^{C\,i} \gamma^\mu \tilde{V}^{b}_{2,\mu} \epsilon^{ab}L_{L}^{j,a} +\textrm{H.c.}
\end{align}
The terms with the third-generation fermions can be written as,
\begin{align}\label{eq:LagV2tp}
\mathcal{L} \supset&\ \tilde{x}^{RL}_{2\ 33}\bar{t}_{R}^{C}\lt\{ -\lt(\gamma\cdot \tilde{V}^{1/3}_{2}\rt) \tau_{L} + \lt(\gamma\cdot \tilde{V}^{-2/3}_{2}\rt) \nu_{L}\rt\}+\textrm{H.c.}
\end{align}

\noindent
$\blacksquare\quad$\underline{$U_{3}$ = ($\mathbf{3}$,$\mathbf{3}$,$2/3$):}\quad
The necessary interaction terms for the triplet $U_3$ are given as,
\ba 
\label{eq:LagU3}
\mathcal{L} \supset x_{3\ ij}^{LL}\bar{Q}_{L}^{i,a}\gamma^{\mu}\lt(\sg^k U_{3,\mu}^k\rt)^{ab}L_{L}^{j,b}  + \textrm{H.c.},
\ea
 where $\sigma^k$ denotes the Pauli matrices. 
The terms for the third-generation fermions can be written explicitly as,
\begin{align}
\mathcal{L} \supset&  x^{LL}_{3\ 33}\Big\{-\bar{b}_{L} \lt(\gamma\cdot U^{2/3}_3\rt) \tau_{L}+\bar{t}_{L} \lt(\gamma\cdot U^{2/3}_3\rt) \nu_{L}\nn\\
&+\sqrt{2}\ \bar{b}_{L} \lt(\gamma\cdot U^{-1/3}_3\rt) \nu_{L}+\sqrt{2}\ \bar{t}_{L} \lt(\gamma\cdot U^{5/3}_3\rt)  \tau_{L}\Big\}\nn\\
&+\textrm{H.c.}\label{eq:LagU3p}
\end{align}

\subsection{Simplified models and benchmark scenarios}\label{subsec:benchmark}
\noindent The models listed above can also be expressed in terms of some simple phenomenological Lagrangians:
\ba 
\label{eq:simpleSlag1}
\mathcal{L} &\supset& \lambda_{\tau}\bigg\{\sqrt{\eta_{L}}\ \bar{t}_{L}^{C}\tau_{L} + \sqrt{\eta_{R}}\ \bar{t}_{R}^{C}\tau_{R}\bigg\}\phi_1 +\lm_{\nu}\bar{b}_{L}^{C}\nu_L\phi_1,\nn\\
&&+{\rm H.c.}, \\
\label{eq:simpleSlag2}
\mathcal{L} &\supset& \hat\lambda_{\tau}\bigg\{\sqrt{\et_L}\ \bar{b}_{R}\tau_{L}+\sqrt{\et_R}\ \bar{b}_{L}\tau_{R}\bigg\}\phi_2 + \hat\lm_{\nu}\bar{t}_{R}\nu_L\phi_2\nn\\&&+{\rm H.c.}, \\
\label{eq:simpleSlag5}
\mathcal{L} &\supset& \tilde\lambda_{\tau}\bigg\{\sqrt{\eta_{L}}\ \bar{t}_{R}\tau_L + \sqrt{\eta_{R}} \bar{t}_{L}\tau_{R}\bigg\}\phi_5+{\rm H.c.},\\
\label{eq:simplevlag1} 
\mathcal{L} &\supset& 
\Lambda_{\tau}\bigg\{\sqrt{\eta_{R}}\ \bar{t}_{L}^{C}\lt(\gamma\cdot\chi_1\rt)\tau_{R} + \sqrt{\eta_{L}}\ \bar{t}_{R}^{C}\lt(\gamma\cdot\chi_1\rt)\tau_{L}\bigg\}\nn\\
&&+~\Lambda_{\n}\ \bar{b}_{R}^{c}\lt(\gamma\cdot\chi_1\rt)\n_L +{\rm H.c.},\\
\label{eq:simplevlag2} 
\mathcal{L} &\supset& 
\hat\Lambda_{\tau}\bigg\{\ep_R\ \sqrt{\eta_{R}}\ \bar{b}_{R}\lt(\gamma\cdot\chi_2\rt)\tau_{R} + \sqrt{\eta_{L}}\ \bar{b}_{L}\lt(\gamma\cdot\chi_2\rt)\tau_{L}\bigg\}\nn\\
&&+~\hat\Lambda_{\n}\ \bar{t}_{L}\lt(\gamma\cdot\chi_2\rt)\n_L +{\rm H.c.},\\
\label{eq:simplevlag5}
\mathcal{L} &\supset& \tilde{\Lambda}_{\tau}\bigg\{\sqrt{\eta_{R}}\  \bar{t}_{R}\lt(\gamma\cdot\chi_5\rt)\tau_{R}+\sqrt{\eta_{L}}\ \bar{t}_{L}\lt(\gamma\cdot\chi_5\rt)\tau_{L}\bigg\}\nn\\
&&+~{\rm H.c.}
\ea
In the simple expressions, a sLQ (vLQ) with charge $=\pm n/3$ is denoted as $\phi_n$ ($\chi_n$). Here, $\eta_{L}$ and $\eta_{R} = 1 -\eta_{L}$ are the charged lepton chirality fractions ~\cite{Mandal:2015vfa,Chandak:2019iwj}. The $\ep_R = \pm1$ in Eq.~\eqref{eq:simplevlag2} is a sign term  to accommodate a possible relative sign between the left-handed and right-handed terms [see Eq.~\eqref{eq:LagV2p}]. We shall consider only real couplings in our analysis for simplicity. 

We illustrate how one can map the standard LQ models to the simplified Lagrangians in Table~\ref{tab:benchmark} with some benchmark scenarios. The scenarios are named according to lepton chiralities following Refs.~\cite{Chandak:2019iwj,Bhaskar:2020gkk}. For example, in the left-coupling (LC) scenarios, the lepton that couples to the LQ along with a top quark is left-handed. Similarly, in the right-coupling (RC) scenarios, it is right-handed. In LCSS (LCOS) scenarios, the LQ couples to a left-handed $\tau$ or a $\nu$ via couplings of equal magnitudes with the same (opposite) sign. In the RLCSS and RLCOS scenarios, the LQ couples with $\n$ and $\ta_R$ with same-sign or opposite-sign couplings. Note, however,  the couplings need not be of the same magnitude in these cases. 
 
 In the table, we also show the decays and branching ratios (BRs) of each LQ assuming it has no other decay modes. In the LC and RC scenarios, the LQ decays to a top quark with $100\%$ BR. In contrast, in the LCSS/OS scenarios, it does so with about $50\%$ BR [except for the vector LCSS scenario where a $U_1$ can decay to a top quark and a $\n$ with less than $50\%$ BR depending on the $x_{1\ 33}^{RR}$ coupling, see Eq.~\eqref{eq:LagU1p}]. For the RLCSS/OS scenarios, we choose the top-quark decay mode BRs to be $50\%$. The discovery/exclusion potential for these LQs mostly depend on the BRs and masses. Hence, scenarios with equal BRs would have similar potentials, except for the scalar LCSS/OS scenarios where the interference between some single production diagrams becomes constructive or destructive depending on the relative sign of the couplings~\cite{Chandak:2019iwj}. In the table, we still show these scenarios to demonstrate how the simple Lagrangians (suitable for a straightforward interpretation of experimental results) can cover all the possibilities.

In the case of vLQs, the kinetic terms of the Lagrangian contains an additional parameter $\kappa$~\cite{Dorsner:2016wpm}.
\begin{align}
\mc L \supset -\frac12\chi^\dag_{\m\n}\chi^{\m\n} + M^2_\chi\ \chi^\dag_\m\chi^\m -ig_s\kp \ \chi^\dag_\m T^a\chi_\n\ G^{a\,\m\n},\label{eq:lkin}
\end{align}
where $\chi_{\m\n}$ stands for the field-strength tensor of $\chi$. We found that the the cross section of the pair and single production processes depends on $\kappa$. Here, we consider two benchmark values, $\kp=0$ and $\kp=1$.


\section{LHC Phenomenology \& Search Strategy}
\label{sec:pheno}
\noindent
We summarise the computational packages used in our analysis. The Lagrangian terms in Eqs.~\eqref{eq:simpleSlag1}--\eqref{eq:simplevlag5} have been encoded in {\sc FeynRules}~\cite{Alloul:2013bka} to generate UFO ~\cite{Degrande:2011ua}
model files. We use {\sc MadGraph5}~\cite{Alwall:2014hca} to generate the signal and background processes at the leading order (LO). 
Only the pair production cross sections of sLQs are corrected with a next-to-leading order (NLO) QCD $K$-factor of 1.3~\cite{Kramer:2004df,Mandal:2015lca}. The higher-order corrections to the SM background processes are known in the literature and their cross sections are corrected using the appropriate $K$-factors. 
We use NNPDF2.3LO~\cite{Ball:2012cx} parton distribution functions  with dynamical renormalisation and factorization scales set  equal to the mass of the LQs. The parton level events are passed through {\sc Pythia6}~\cite{Sjostrand:2006za} for showering and hadronisation. We then use {\sc Delphes3}~\cite{deFavereau:2013fsa} for simulating the detector effects with the default CMS card (we only modify the $b$-tagging efficiency to set it at a flat $85\%$.) We reconstruct  fatjets from {\sc Delphes} tower objects using the anti-$k_t$~\cite{Cacciari:2008gp} clustering algorithm (with $R = 0.8$) in {\sc FastJet}~\cite{Cacciari:2011ma}. 

\subsection{Production at the LHC}

\begin{figure*}[!t]
\captionsetup[subfigure]{labelformat=empty}
\subfloat[\quad\quad\quad(a)]{\includegraphics[width=0.9\columnwidth, height=4.5cm]{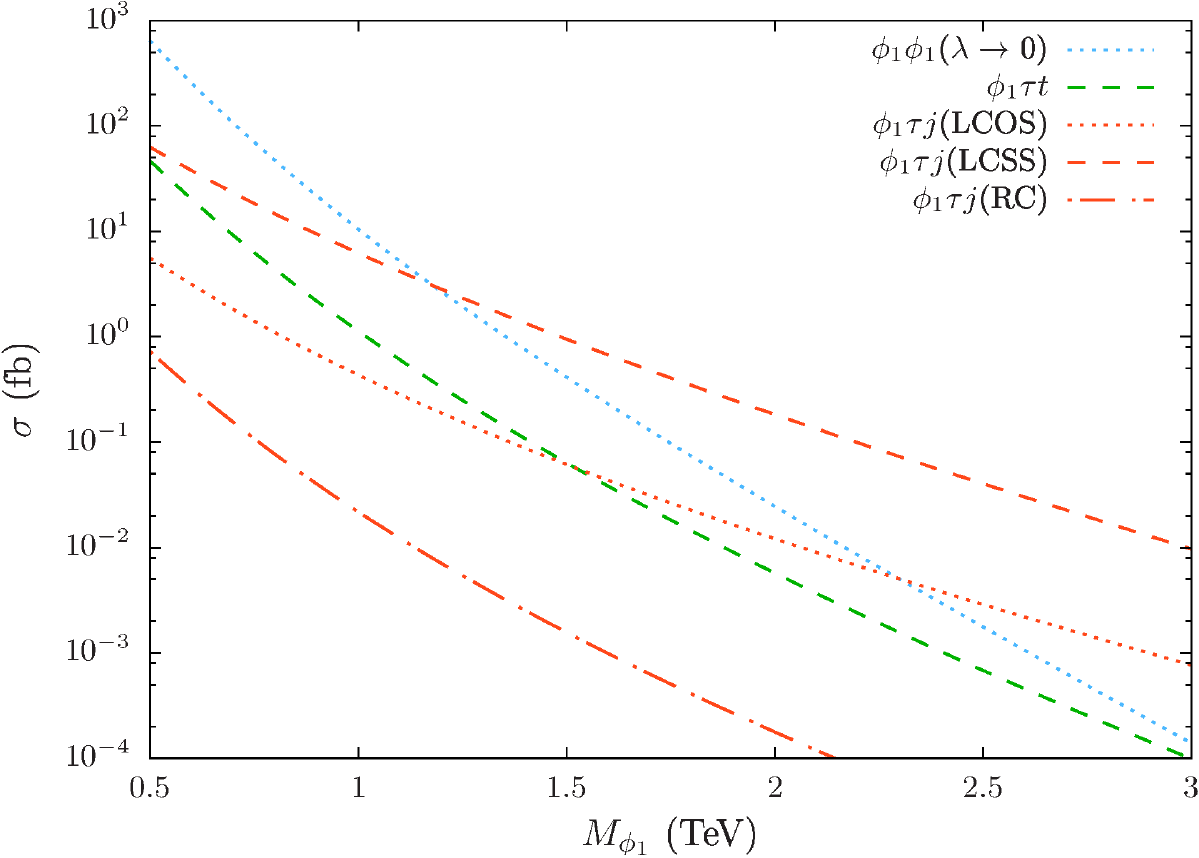}\label{fig:xsec01}}\quad\quad
\subfloat[\quad\quad\quad(b)]{\includegraphics[width=0.9\columnwidth, height=4.5cm]{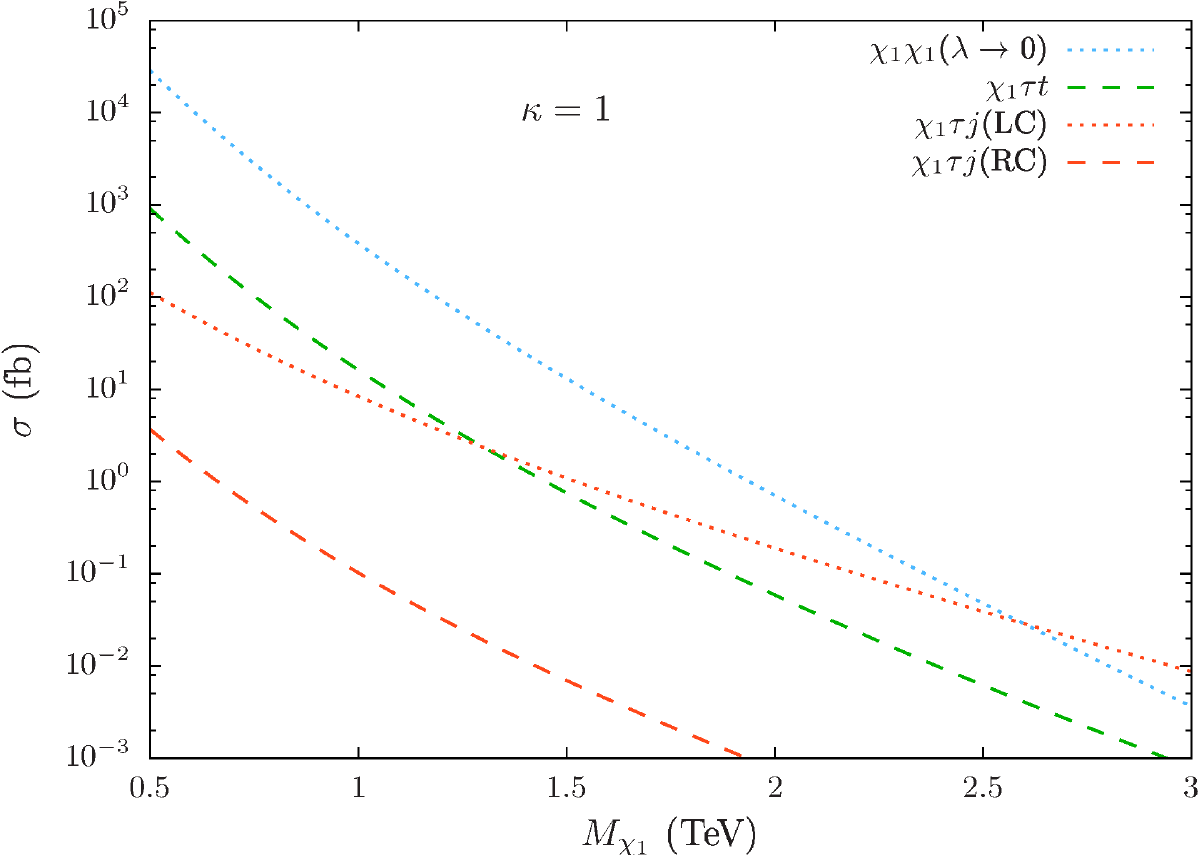}\label{fig:xsec02}}\\
\subfloat[\quad\quad\quad(c)]{\includegraphics[width=0.9\columnwidth, height=4.5cm]{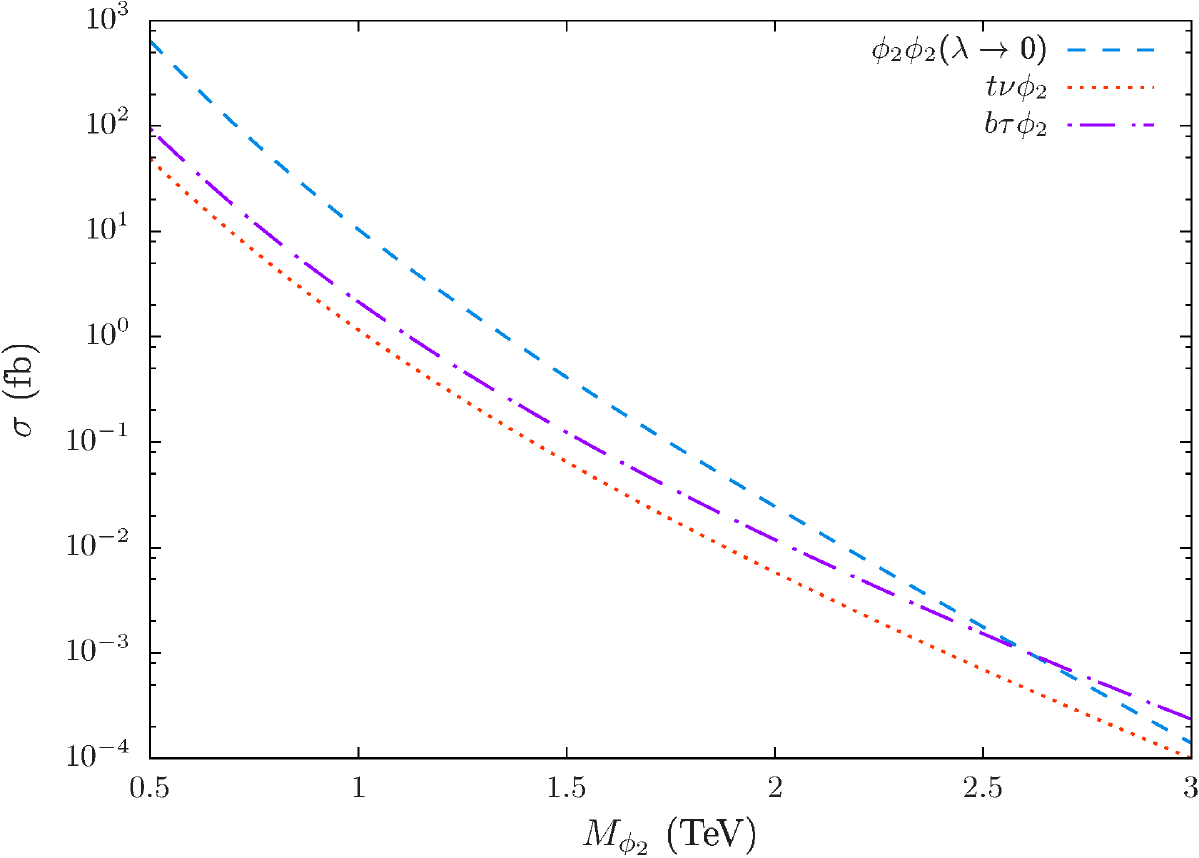}\label{fig:xsec03}}\quad\quad
\subfloat[\quad\quad\quad(d)]{\includegraphics[width=0.9\columnwidth, height=4.5cm]{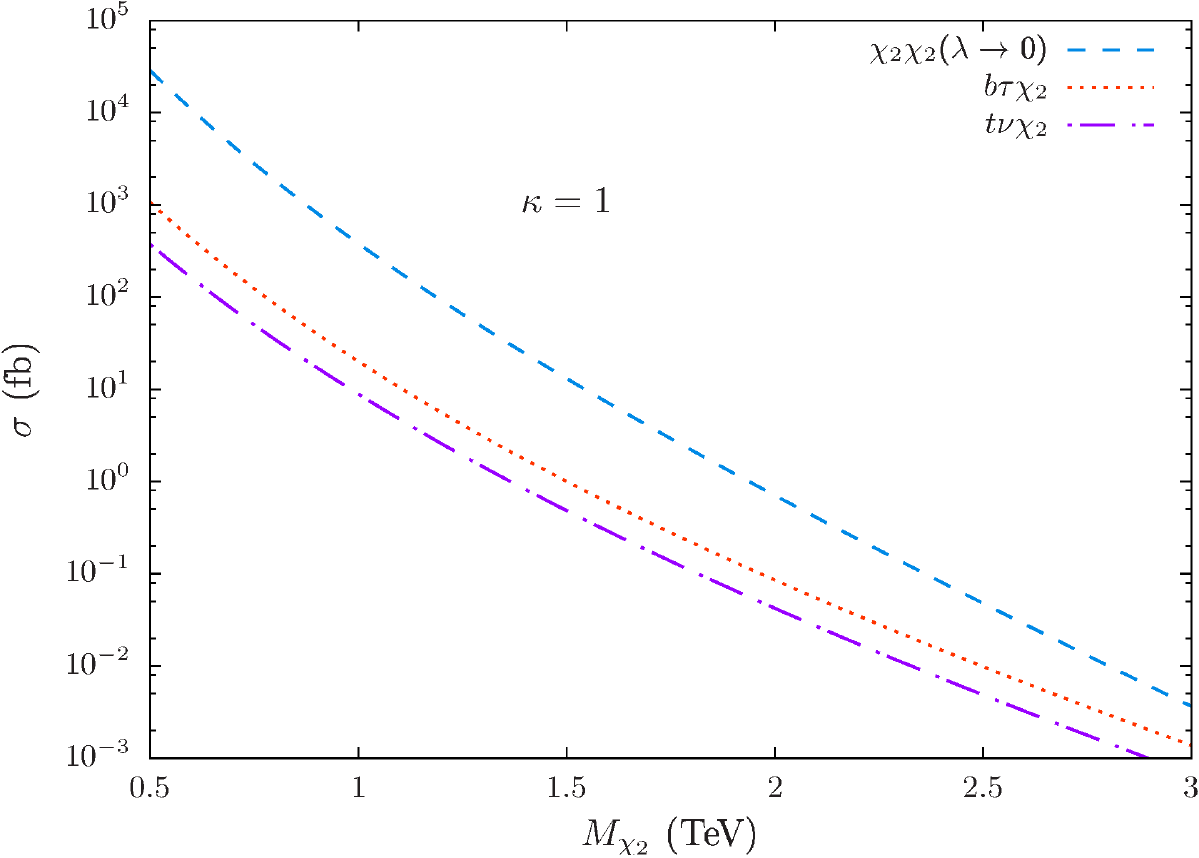}\label{fig:xsec04}}\\
\subfloat[\quad\quad\quad(e)]{\includegraphics[width=0.9\columnwidth, height=4.5cm]{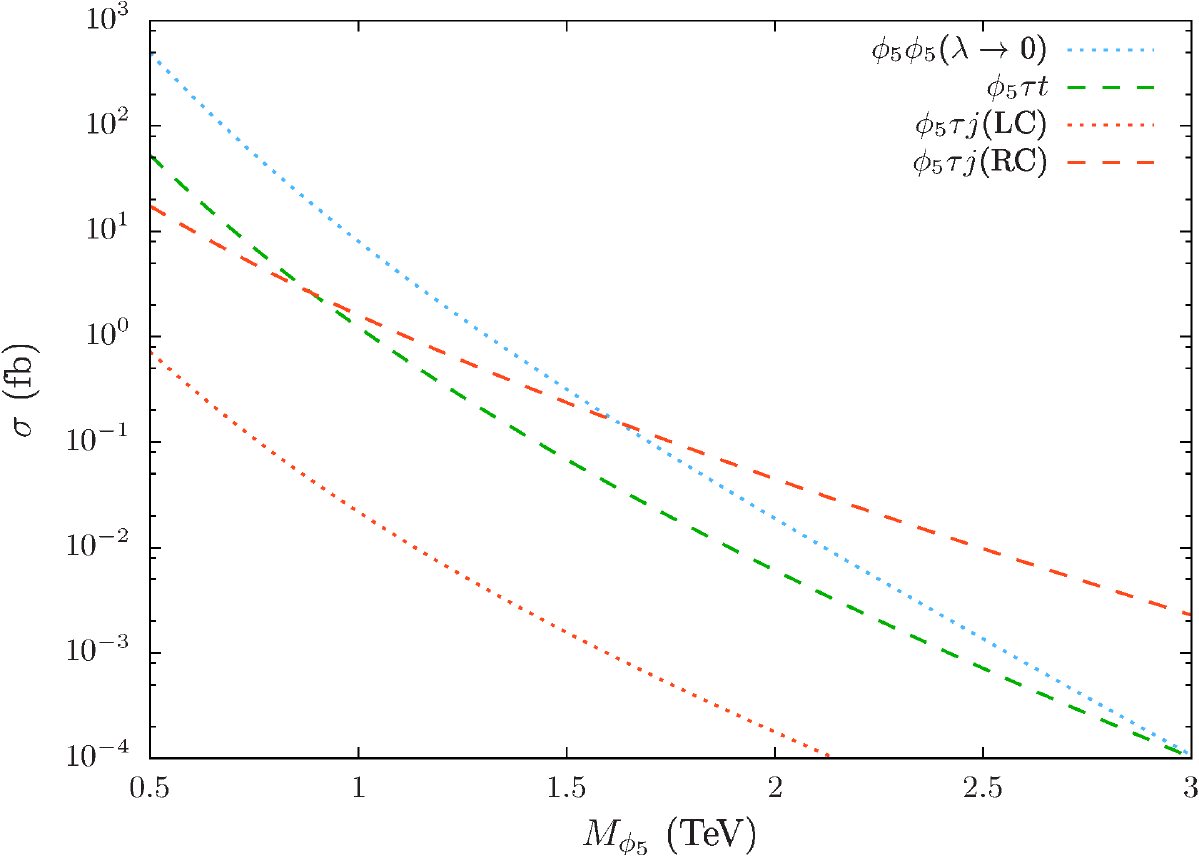}\label{fig:xsec05}}\quad\quad
\subfloat[\quad\quad\quad(f)]{\includegraphics[width=0.9\columnwidth, height=4.5cm]{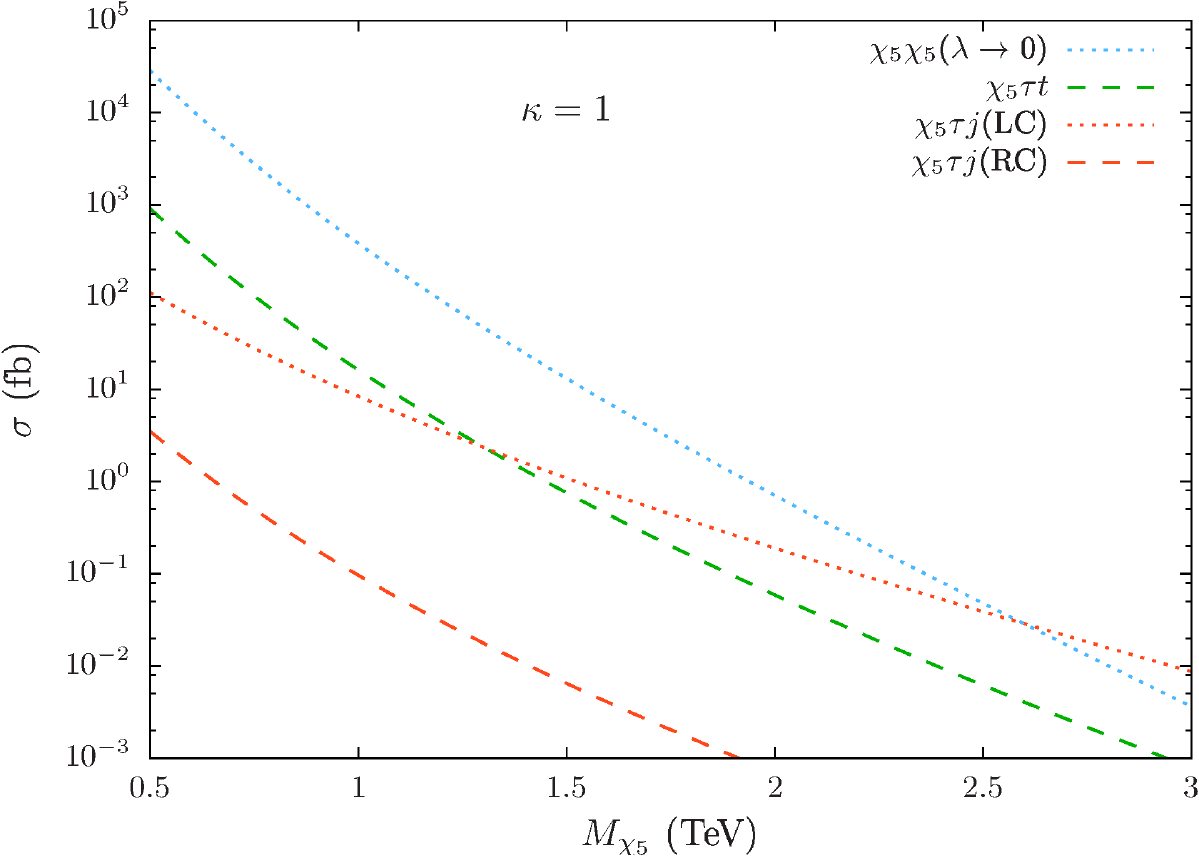}\label{fig:xsec06}}\\
\caption{The parton-level cross sections of different production channels of $\phi_i$ [(a), (c), (e)]/$\chi_i$ [(b), (d), (f)] at the $14$ TeV LHC as functions of $M_{\chi_n/\phi_n}$. The single production cross sections are computed for a benchmark coupling 
$\lm, \Lm = 1$ (see Table~\ref{tab:benchmark}). For the VLQs, we have shown the scenarios with QCD coupling $\kappa=1$.  Here, we consider $\tau j$ and $\tau t$ in the single production processes and the $j$ includes all the light jets including $b$-jets. Their cross sections are generated with a cut on the transverse momentum of the jet, $p^j_{\rm T} > 20$ GeV.}
\label{fig:xsec}
\end{figure*}

\noindent
Both the sLQs and vLQs can be produced resonantly at the LHC via the pair production and the single production processes. The pair production processes are almost independent of $\lm$; they mostly depend on the universal QCD coupling. 
In the case of vLQs, the free parameter $\kappa$ from Eq.~\eqref{eq:lkin}, can affect both pair and single production cross sections. The pair production channels can have the following final states,
\begin{eqnarray}
pp\ \to\left\{\begin{array}{ccl}
 \phi_1\phi_1/\chi_1\chi_1&\to& (t\tau)(t\tau)/(t\tau)(b\n)/(b\n)(b\n)\\
 \phi_2\phi_2/\chi_2\chi_2&\to& (t\nu)(t\nu)/(t\nu)(b\tau)/ (b\tau)(b\tau)\\
 \phi_5\phi_5/\chi_5\chi_5&\to& (t\tau)(t\tau)
\end{array}\right\}.\label{eq:pair_all}\nn\\
\end{eqnarray}

In our previous papers~\cite{Chandak:2019iwj,Bhaskar:2020gkk}, we considered the symmetric pair production channels where both the LQs decay to the same final state and considered single productions yielding similar final states. In this paper, we also explore the asymmetric pair production channels with
$(t\tau)(b\nu)$ or $(t\nu)(b\tau)$ final states while ignoring 
final states without a top quark or a $\tau$.
We include single production channels giving  similar final states (we refer to these as the asymmetric single production channels).
Though it is generally assumed that the symmetric modes are  more promising for discovery~\cite{Diaz:2017lit}, due to combinatorics, the asymmetric modes can also provide comparable (or, in some cases, even better) reach. 
From Table~\ref{tab:benchmark}, we see that the charge-$1/3$ or charge-$5/3$ LQs can have the $(t\tau)(t\tau)$ mode, whereas and $tb\tau\nu$ can come from charge-$1/3$ or charge-$2/3$ LQs.

A three-body single production, where a LQ is produced in association with a $\tau$ and either a jet or a top quark, can produce similar final states as the pair production. This allows us to combine pair and single productions in the same analysis and enhance the discovery reach~\cite{Mandal:2012rx,Mandal:2015vfa,Mandal:2016csb,Chandak:2019iwj,Bhaskar:2020gkk}. We show the possible symmetric and asymmetric  single production modes below. 
\begin{eqnarray}
pp&\ \to&\left\{\begin{array}{ccl}
\phi_1/\chi_1\ t \tau &\to & (t\tau)t\tau / (b\n)t\tau\\
\phi_1/\chi_1\ b\n &\to & (t\tau)b \n/(b\n)b \n\\
\phi_1/\chi_1\ \tau j &\to &  (t\tau)\tau j /(b\n)\tau j 
\end{array}\right\},\label{eq:single1}
\end{eqnarray}\begin{eqnarray}
pp&\ \to&\left\{\begin{array}{ccl}
\phi_2/\chi_2\ t \nu &\to & (t\nu)t\nu/(b \tau)t \nu \\
\phi_2/\chi_2\ b \tau &\to &  (t \nu)b \tau/(b \tau)b \tau\\
\phi_2/\chi_2\ \nu j &\to & (t \nu)\nu j/(b\tau)\nu j \\
\end{array}\right\},\label{eq:single2}\\
pp&\ \to&\left\{\begin{array}{ccl}
\phi_5/\chi_5\ t \tau &\to & (t \tau)t \tau \\
\phi_5/\chi_5\ \tau j &\to &  (t \tau)\tau j
\end{array}\right\}.\label{eq:single5}
\end{eqnarray}
Of these, we ignore the modes without at least one top quark and/or a $\ta$.
In principle, there are two $bg$ initiated two-body single production processes also where a LQ is produced in association with a $\tau$ or a $\n$. We ignore them to avoid double counting with the $gg$ initiated three-body processes where a gluon splits into a pair of $b$ quarks. This is justified since the $b$ quark has to be hard in order to produce a TeV-scale LQ.

In Fig.~\ref{fig:xsec}, we show the parton-level cross sections of different production processes of
scalar (left column) and vector (right column) LQs as functions of their masses. We compute the cross sections with $\lm=1$ for the sLQs, and $\Lm=1$ and $\kp=1$ for the vLQs. From these figures, we get an idea about the relative magnitudes of the cross sections and the  relative importance of processes (\emph{a priori} to the cuts) in the mass range of our interest. In Fig.~\ref{fig:xsec01}, two types of LC scenarios, namely the LCOS and LCSS scenarios  are separately shown for the $pp\to\phi_1\tau j$ single production process. Both scenarios have the same BRs of $\phi_1$, i.e., $50$\% each in $t\tau$ and $b\nu$ modes. But the cross section for the LCSS scenario is bigger than that in the LCOS scenario. This is because, the $\lm_{\ell}=\lm_{\nu}$ relation in the LCSS scenario results in a constructive interference between some of the single production diagrams which becomes destructive in the LCOS scenario where $\lm_{\ell}= - \lm_{\nu}$. Such difference is not observed for other LQ species.

Single production cross sections vary as the square of the new coupling $\lm$ or $\Lm$. In the figure all processes have been computed for $\lm=\Lm=1$. With order one $\lm$ (or $\Lm$), it is possible for some single production modes to have bigger cross sections than the pair production in the mass range of our interest.
For example, we see that in the LC scenario, $pp\to \chi_1\tau j$  overtakes the pair production processes at $2.6$ TeV for $\kappa=1$ [Fig.~\ref{fig:xsec02}] but the cross section for the similar $\chi_1\tau t$ process remains 
lower. Similar behaviour can be seen for the $pp\to \phi_1\tau j$ process also. However, in some scenarios, these processes have much smaller cross section. For example, in the RC scenario, the $pp\to \phi_1\tau j, \chi_1\tau j, \chi_5\tau j$ processes has much smaller cross section than the rests in the entire mass range considered. Similarly, $pp\to \phi_5\tau j$
have small cross section in the LC scenario. This is mainly because, in these scenarios, the LQs are produced from a right-handed top quark generated in the charge current interaction via a chirality flip.
In the figure, we just show the plots for the vLQs with $\kp=1$. If $\kp=0$, the pair production cross sections become smaller and hence the crossover points appear at lower masses.

\subsection{Signal topologies and the background}
\noindent From Eqs.~\eqref{eq:pair_all}-\eqref{eq:single5} we see that if we disregard the final states without a single top quark or a $\ta$, the remaining ones can be identified as of two types---one with one top quark and two $\ta$'s and the other with one top quark and only one $\ta$. As before, we will consider only hadronically-decaying top quarks in our analysis to exploit its boosted nature.
A $\tau$ can decay either hadronically or leptonically with branching fractions of about $65$\% and $35$\%, respectively. This gives three possible final states in terms of $\tau$ decays---$\tau_h\tau_h$, when both of them decay hadronically with a probability about $42$\%, $\tau_h\tau_\ell$, when one of them decays hadronically and the other leptonically with about $46$\% probability, and $\tau_\ell\tau_\ell$, when both of them decay leptonically with a probability of only $12$\%. In our analysis, we consider both hadronic and semileptonic $\tau$'s and study the following signatures:
\begin{enumerate}
    \item[A.] at least one hadronically-decaying top quark along with either two hadronically-decaying $\tau$ leptons ($t_h\ta_h\ta_h$) or a hadronically-decaying $\tau$ and a leptonically-decaying one ($t_h\ta_h\ta_{\ell}$),
    \item[B.] at least one hadronically-decaying top quark with only one $\ta$ decaying hadronically and some missing energy ($t_h\ta_h$+MET).
\end{enumerate}
The first signature would capture the symmetric final states, while the second one, the asymmetric ones. The other symmetric final state with a pair of top quarks leading to the $t_ht_h$+MET signature ($t\n t\n$) has been studied in Ref.~\cite{Vignaroli:2018lpq}. 
As already discussed, the above signal topologies can arise from both pair and single production processes. For instance, the signal $t_h\tau_h\tau_h$ can come from $pp\to \chi_{1,5}\chi_{1,5}$~(or $pp\to \phi_{1,5}\phi_{1,5}$) and $pp\to \chi_{1,5}t\tau$~(or $pp\to \phi_{1,5}t\tau$) processes. If not carefully done, this might lead to double counting while generating the signal processes. To circumvent this issue, we ensure that the LQ and its antiparticle are never on-shell simultaneously while generating single production events~\cite{Mandal:2015vfa}.

\begin{table}[t!]
\centering{\linespread{2}
\begin{tabular*}{\columnwidth}{l @{\extracolsep{\fill}} crc }
\hline
\multicolumn{2}{l}{Background } & $\sg$ & QCD\\ 
\multicolumn{2}{l}{processes}&(pb)&order\\\hline\hline
\multirow{2}{*}{$V +$ jets~ \cite{Catani:2009sm,Balossini:2009sa}  } & $Z +$ jets  &  $6.33 \times 10^4$& NNLO \\ \cline{2-4} 
                & $W +$ jets  & $1.95 \times 10^5$& NLO \\ \hline
\multirow{3}{*}{$VV +$ jets~\cite{Campbell:2011bn}}   & $WW +$ jets  & $124.31$& NLO\\ \cline{2-4} 
                  & $WZ +$ jets  & $51.82$ & NLO\\ \cline{2-4} 
                   & $ZZ +$ jets  &  $17.72$ & NLO\\ \cline{1-4}
\multirow{3}{*}{Single $t$~\cite{Kidonakis:2015nna}}  & $tW$  &  $83.10$ & N$^2$LO \\ \cline{2-4} 
                   & $tb$  & $248.00$ & N$^2$LO\\ \cline{2-4} 
                   & $tj$  &  $12.35$ & N$^2$LO\\  \cline{1-4}
$tt$~\cite{Muselli:2015kba}  & $tt +$ jets  & $988.57$ & N$^3$LO\\ \cline{1-4}
\multirow{2}{*}{$ttV$~\cite{Kulesza:2018tqz}} & $ttZ$  &  $1.05$ &NLO+NNLL \\ \cline{2-4} 
                   & $ttW$  & $0.65$& NLO+NNLL \\ \hline
\end{tabular*}}
\caption{Total cross sections without any cut for the SM background processes considered in our analysis. The higher-order QCD cross sections are taken from the literature and the corresponding orders are shown in the last column. We use these cross sections to compute the $K$ factors which we multiply with the LO cross sections to include higher-order effects.}
\label{tab:Backgrounds}
\end{table}

\begin{table*}[]
\centering{\linespread{2}
\begin{tabular*}{\textwidth}{l @{\extracolsep{\fill}}ll}
\hline \multirow{2}{*}{Steps} & \multicolumn{2}{c}{Criteria}\\\cline{2-3}
 & $2\ta$ ($t_h\tau_h\tau_h + t_h\tau_h\tau_\ell$)& $1\ta$ ($t_h\tau_h$+MET) \\ \hline\hline
Leptons and jets basic selection & \begin{tabular}[c]{@{}l@{}}$p_{\rm T}(\ell)>100$ GeV, $|\eta_{\ell}|<2.5$\\ (excluding $1.37<|\eta_e|<1.52$)\\ \hline  $p{\rm _T}(j)>30$ GeV, $|\eta_j|<5.0$,\\$p_{\rm T}(\tau_h)>150$ GeV\end{tabular} & \begin{tabular}[c]{@{}l@{}}p$_T(\ell)>100$ GeV $|\eta_{\ell}|<2.5$\\ (excluding $1.37<|\eta_e|<1.52$)\\\hline p$_T(j)>30$ GeV,  $|\eta_j|<5.0$,\\p$_T(\tau_h)>250$ GeV\end{tabular} \\ \hline
\begin{tabular}[c]{@{}l@{}}Number of leptons/jets,\\ Mass/energy cuts\end{tabular} & \begin{tabular}[c]{@{}l@{}}
$N(\tau_h)=2$ or $N(\tau_h)=N(\ell)=1$,\\ $N(b)>0$\\ \hline $M(\tau_{h_1},\tau_{h_2})>250$ GeV or\\ $M_{\rm T}(\tau_h,\ell,\slashed E_{\rm T})>400$ GeV\end{tabular} & \begin{tabular}[c]{@{}l@{}}
$N(\tau_h)=1$,\\ $N(b)>0$\\\hline $\slashed E_{\rm T}>300$ GeV,\\ $M_{\rm T}(\tau_h,\slashed E_{\rm T})>300$ GeV\end{tabular} \\ \hline
At least one high $p_{\rm T}$ lepton & $p_{\rm T}(\hat\ell)>250$ GeV $(\hat\ell = e,\mu,\ta_h)$& -- \\ \hline
\begin{tabular}[c]{@{}l@{}}At least one top quark\\($t$ identification)\end{tabular} & \begin{tabular}[c]{@{}l@{}}$p_{\rm T}({\rm fatjet})>200$ GeV,\\ $\Delta R({\rm fatjet},\hat\ell_1)$, $\Delta R({\rm fatjet},\hat\ell_2)>0.8$, \\ $\tau_{32}<0.81$, $\tau_{21}>0.35$,\\\hline $M({\rm fatjet})>120$ GeV,\\ $S_{\rm T}>1300$ GeV\end{tabular} & \begin{tabular}[c]{@{}l@{}}$p_{\rm T}({\rm fatjet})>200$ GeV,\\ $\Delta R({\rm fatjet},\tau_h)>0.8$\\ $\tau_{32}<0.81$, $\tau_{21}>0.35$,\\\hline $M({\rm fatjet})>120$ GeV,\\ $S_{\rm T}>1100$ GeV\end{tabular} \\ \hline
\end{tabular*}}
\caption{The sets of cuts applied for the two signatures. The cuts are motivated by the CMS analysis with $137$ fb$^{-1}$~\cite{Sirunyan:2020zbk}.}
\label{tab:cuts}
\end{table*}

\begin{table*}[]
\centering{\linespread{2}
\begin{tabular*}{\textwidth}{c @{\extracolsep{\fill}}||c c c | c c | c c | c || c | c | c | c }
\hline
 \multirow{3}{*}{~\rotatebox[origin=c]{90}{Significance $\mc Z$ }} & \multicolumn{11}{c}{Limit on $M_\phi$ (TeV)}\\ 
	& \multicolumn{8}{c}{$t_h\tau_h\tau_h+t_h\tau_h\tau_\ell$}	& \multicolumn{4}{c}{$t_h\tau_h$+MET} \\ \cline{2-13} 
                         & \multicolumn{5}{c|}{$\phi_1$} & \multicolumn{3}{c||}{$\phi_5$} & \multicolumn{2}{c|}{$\phi_1$} & \multicolumn{2}{c}{$\phi_2$} \\ 
\cline{2-13}
&\multicolumn{3}{c|}{Combined}&\multicolumn{2}{c|}{Pair}&\multicolumn{2}{c|}{Combined}&Pair&\multicolumn{1}{c|}{Combined}&\multicolumn{1}{c|}{Pair}&\multicolumn{1}{c|}{Combined}&\multicolumn{1}{c}{Pair}\\
\cline{2-13}
& LCOS & LCSS & RC & BR=$0.5$ & BR=$1$ & LC & RC & BR=$1$ & LCSS & BR=$0.5$ & RLCSS & BR=$0.5$ \\
   \hline\hline
~~5 & 0.99 & 1.07 & 1.33 & 0.96 & 1.31 & 1.33 & 1.34 & 1.31 & 1.16 & 1.10 & 1.13 & 1.09  \\ \hline
~~3 & 1.13 & 1.23 & 1.44 & 1.10 & 1.23 & 1.44 & 1.45 & 1.42 & 1.32 & 1.24 & 1.27 & 1.23 \\ \hline
~~2 & 1.23 & 1.36 & 1.52 & 1.19 & 1.50 & 1.53 & 1.53 & 1.51 & 1.42 & 1.34 & 1.37 & 1.33 \\ \hline
\end{tabular*}}
\caption{The mass limits corresponding to $5\sg$ (discovery), $3\sg$ and $2\sg$ (exclusion) significances ($\mc Z$) for observing the $\phi_1$, $\phi_2$ and $\phi_5$ signals over the SM backgrounds for 3 ab$^{-1}$ integrated luminosity at the $14$ TeV LHC with combined and pair-production-only signals. Here, we show the mass limits for the sLQs for both the signatures described in the paper.}
\label{tab:sig2}
\medskip


\centering{\linespread{2}
\begin{tabular*}{\textwidth}{c ||@{\extracolsep{\fill}} c c c c | c c | c c | c || c c c c | c c | c c | c }
\hline
 \multirow{3}{*}{~\rotatebox[origin=c]{90}{Significance $\mc Z$ }} & \multicolumn{18}{c}{Limit on $M_\chi$ (TeV) (Signature:~$t_h\tau_h\tau_h+t_h\tau_h\tau_\ell$)}\\ 
	& \multicolumn{8}{c}{$\kappa=0$}	& \multicolumn{10}{c}{$\kappa=1$} \\ \cline{2-19} 
                         & \multicolumn{6}{c|}{$\chi_1$} & \multicolumn{3}{c||}{$\chi_5$} & \multicolumn{6}{c|}{$\chi_1$} & \multicolumn{3}{c}{$\chi_5$} \\ 
\cline{2-19}
&\multicolumn{4}{c|}{Combined}&\multicolumn{2}{c|}{Pair}&\multicolumn{2}{c|}{Combined}&Pair&\multicolumn{4}{c|}{Combined}&\multicolumn{2}{c|}{Pair}&\multicolumn{2}{c}{Combined}&Pair\\
\cline{2-19}
& LC50 & LC & RC50 & RC & BR=$0.5$ & BR=$1$ & LC & RC & BR=$1$ & LC50 & LC & RC50 & RC & BR=$0.5$ & BR=$1$ & LC & RC & BR=$1$\\
   \hline\hline
~~5 & 1.49 & 1.75 & 1.43 & 1.69 & 1.41 & 1.68 & 1.75 & 1.69 & 1.68 & 1.78 & 2.05 & 1.76 & 2.03 & 1.74 & 2.02 & 2.05 & 2.03 & 2.01 \\ \hline
~~3 & 1.60 & 1.87 & 1.53 & 1.80 & 1.51 & 1.78 & 1.87 & 1.80 & 1.78 & 1.90 & 2.16 & 1.87 & 2.13 & 1.85 & 2.12 & 2.16 & 2.13 & 2.12 \\ \hline
~~2 & 1.69 & 1.96 & 1.61 & 1.88 & 1.59 & 1.86 & 1.96 & 1.88 & 1.86 & 1.98 & 2.25 & 1.95 & 2.21 & 1.92 & 2.20 & 2.25 & 2.21 & 2.20 \\ \hline
\end{tabular*}}
\caption{Same as Table~\ref{tab:sig2} for $\chi_1$ and $\chi_5$. The LC50 and RC50 represent the 
cases where the BR of $\chi_{1}\to t\tau$ mode is $50$\% and the LC (RC) stands for 100\% BR case. We have shown the mass limits for $\kappa=0$ and $1$. The signature which we consider for these mass limits is $t_h\tau_h\tau_h + t_h\tau_h\tau_\ell$.}
\label{tab:sig1}
\medskip


\centering{\linespread{2}
\begin{tabular*}{\textwidth}{c @{\extracolsep{\fill}}|| c | c | c | c || c | c | c | c }
\hline
 \multirow{3}{*}{~\rotatebox[origin=c]{90}{Significance $\mc Z$ }} & \multicolumn{8}{c}{Limit on $M_\chi$ (TeV) (Signature:~$t_h\tau_h+$MET)}\\ 
	& \multicolumn{4}{c}{$\kappa=0$}	& \multicolumn{4}{c}{$\kappa=1$} \\ \cline{2-9} 
                         & \multicolumn{2}{c|}{$\chi_1$} & \multicolumn{2}{c||}{$\chi_2$} & \multicolumn{2}{c|}{$\chi_1$} & \multicolumn{2}{c}{$\chi_2$} \\ 
\cline{2-9}
&\multicolumn{1}{c|}{Combined}&\multicolumn{1}{c|}{Pair}&\multicolumn{1}{c|}{Combined}&Pair&\multicolumn{1}{c|}{Combined}&\multicolumn{1}{c|}{Pair}&\multicolumn{1}{c}{Combined}&Pair\\
\cline{2-9}
& RLC50 & BR=$0.5$ & LCSS & BR=$0.5$ & RLC50 & BR=$0.5$ & LC50 & BR=$0.5$\\
   \hline\hline
~~5 & 1.58 & 1.53 & 1.56 & 1.53 & 1.90 & 1.88 & 1.88 & 1.87 \\ \hline
~~3 & 1.69 & 1.64 & 1.66 & 1.63 & 2.01 & 1.98 & 1.99 & 1.97 \\ \hline
~~2 & 1.79 & 1.72 & 1.74 & 1.71 & 2.10 & 2.07 & 2.08 & 2.06 \\ \hline
\end{tabular*}}
\caption{Same as Table~\ref{tab:sig2} for $\chi_1$ and $\chi_5$. Here, we show the mass limits for vLQs for the $t_h\tau_h+\textrm{MET}$ signature. The RLC50 represents RLCSS/OS scenarios.}
\label{tab:sig3}
\end{table*}

The SM background of our signatures is very large and requires carefully designed kinematic cuts to make the the signal observable. The SM background for both $(t_h\tau_h\tau_h+t_h\tau_h\tau_\ell)$ and $(t_h\tau_h+\textrm{MET})$ signals would contain at least one toplike fatjet. In the SM, it can arise directly from a top quark decaying hadronically or indirectly when a bunch of QCD jets combine to mimic the features of a toplike fatjet. In addition, the SM background should also contain $\tau$-tagged high-$p_{\rm T}$ jets (from hadronic $\tau$ decays) and high-$p_{\rm T}$ light leptons (since there is a leptonic $\tau$ decay in the signal). Although small, a QCD jet can sometime appear as a $\tau_h$ due to mistagging. Since some background processes have huge cross sections, even small mistagging rates might lead to a substantial number of background events at the end.
The relevant background processes for the two signatures are listed in Table~\ref{tab:Backgrounds} where the available highest-order values of their total cross sections are shown. We generate all the background processes at LO and scale the cross sections with the appropriate $K$-factors. 
Notice that some of the background processes have very high cross sections. For these processes, we apply generation-level cuts to save computation time and improve the statistics.
We briefly discuss the background processes below.

For both signal topologies, the dominant background is from the inclusive $t\bar{t}+\textrm{jets}$ process. All three $t\bar{t}$ decay modes, i.e., hadronic, semileptonic and leptonic contribute to the background---the semileptonic mode contributes dominantly followed by the leptonic and the hadronic modes. For one-$\tau$ signature, the second dominant background is $W+\textrm{jets}$ whereas for the two-$\tau$ category, it is $tW$. We have included other relevant but subdominant background processes like $V+\textrm{jets}$, $VV+\textrm{jets}$, $t+\textrm{jets}$, and $ttV$, etc., in our analysis.



\subsection{Analysis}
\noindent
For both signal topologies, we use the anti-$k_t$ clustering algorithm and make use of two different levels of jet information---we
use two types of jets with different values of the jet radius parameter $R$.  We call the jets with $R=0.4$  ``AK4-jets'' and the $R=0.8$ ones as ``AK8-fatjets''. We use the Delphes tower objects to construct the AK8-fatjets and identify the hadronic top from them with the following criteria:
\begin{itemize}
    \item[--] Mass of the AK8-fatjet, $M_{fj}>120$~GeV and its transverse momentum $p_{\rm T}>200$~GeV.
    \item[--] The subjettiness ratios $\tau_{32}<0.81$ and $\tau_{21}>0.35$ where $\tau_{ij}\equiv \tau_{i}/\tau_{j}$.
\end{itemize}
The subjettiness criteria is motivated by the fact that $\tau_N$ assumes smaller values as the fatjet closely resembles a collection of $N$ subjets. The AK4-jets are used to identify the $\tau_h$- and $b$-tagged jets. 

We list the  selection cuts  used for the two signal categories sequentially (cuts-flow) in Table~\ref{tab:cuts}.  We use cuts on the invariant mass and the transverse mass $M_{\rm T}$ of the leptons. The transverse mass is defined as
\begin{eqnarray}
 M^2_{\rm T}(A,B) &=& 2p_{\rm T}^Ap_{\rm T}^B \Big\{1 - \cos\Delta\phi(p_{\rm T}^{A},p_{\rm T}^{B})\Big\},\quad\\
 M^2_{\rm T}(A,B,\slashed E_{\rm T})&=&
  M_{\rm T}^2(A,B)+M_{\rm T}^2(A,\slashed E_{\rm T})\nonumber\\
  && + M_{\rm T}^2(B,\slashed E_{\rm T}).
\end{eqnarray}
To make sure that there is no overlap between a $\tau_h$ and the hadronic top, we demand a radial separation between the fatjet and a $\tau_h$ while identifying the top. We also apply a cut on ${\rm S_T}$,  the scalar sum of visible momenta.
Notice that for both signatures, we demand the presence of at least one $b$-tagged AK4-jet. This reduces the $V+$ jets backgrounds significantly (in addition to the invariant mass cut on the $\tau_h$ pair in the $2\ta_h$ signature to avoid the $Z+$ jets background). This is justified since at least one top quark should be always present in the final states. The $b$-jet criterion, however, is completely inclusive in nature and we do not use the $b$-jet(s) for any reconstruction.

\section{HL-LHC prospects}\label{sec:dispot}

\begin{figure*}[]
\centering
\captionsetup[subfigure]{labelformat=empty}
\subfloat[\quad\quad\quad(a)]{\includegraphics[width=0.9\columnwidth, height=4.5cm]{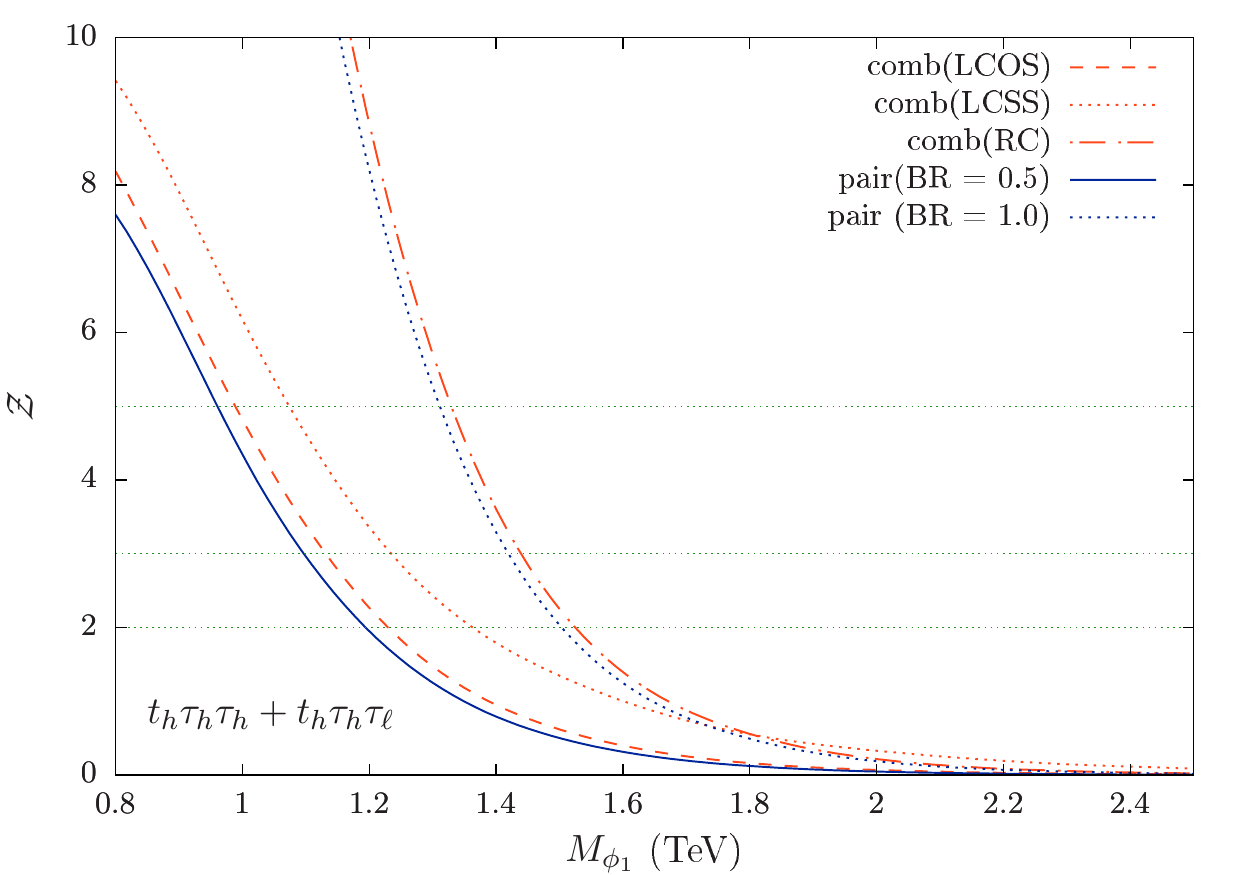}\label{fig:xset101}}\quad\quad\quad\quad
\subfloat[\quad\quad\quad(b)]{\includegraphics[width=0.9\columnwidth, height=4.5cm]{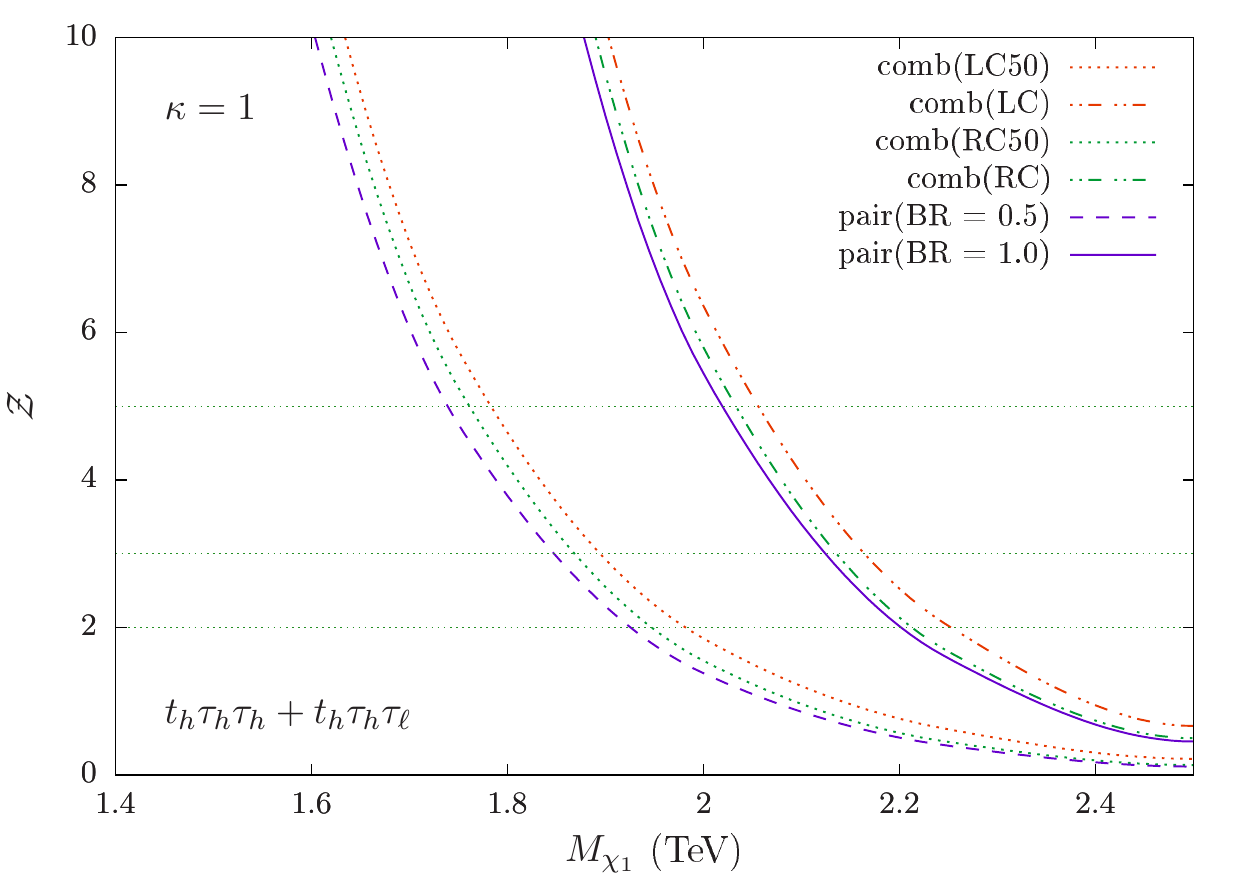}\label{fig:xset102}}\\
\subfloat[\quad\quad\quad(c)]{\includegraphics[width=0.9\columnwidth, height=4.5cm]{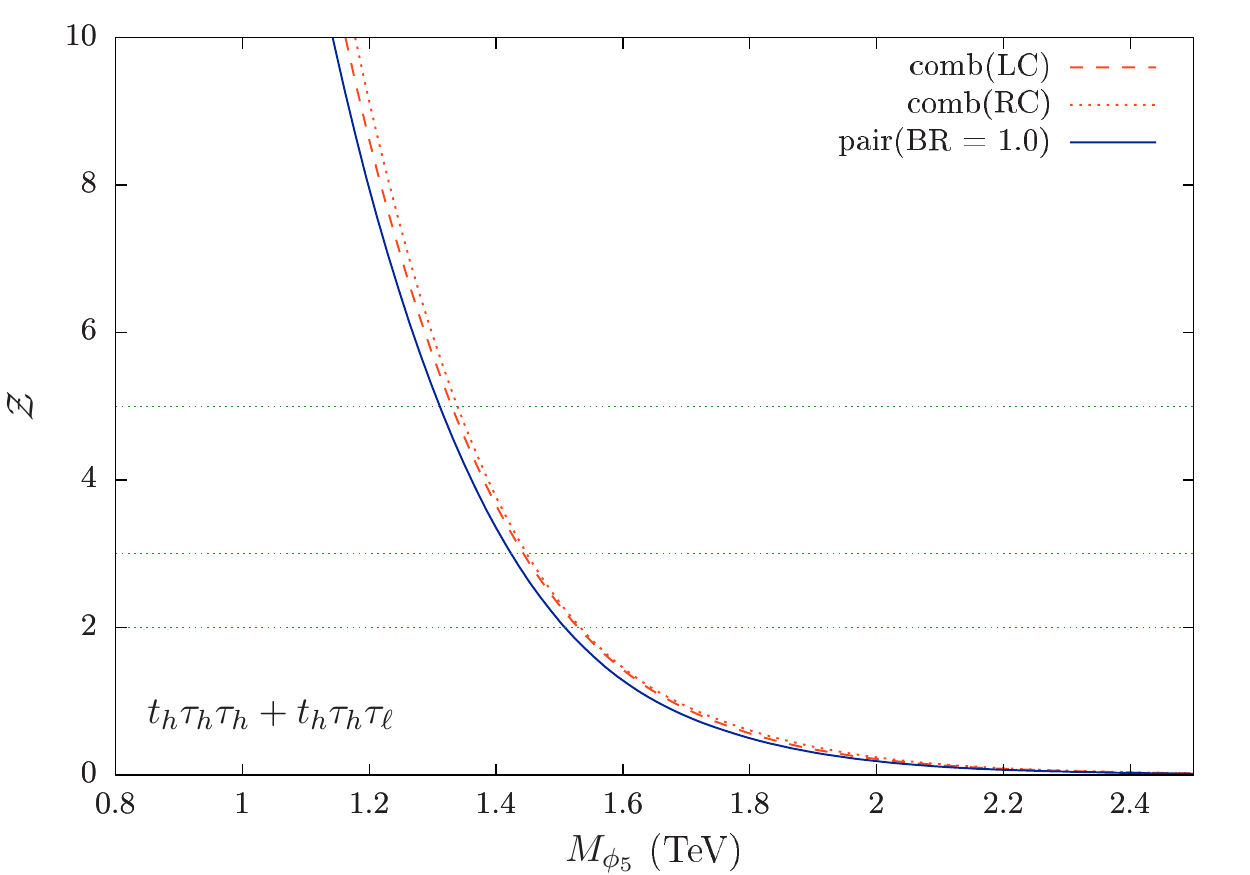}\label{fig:xset103}}\quad\quad\quad\quad
\subfloat[\quad\quad\quad(d)]{\includegraphics[width=0.9\columnwidth, height=4.5cm]{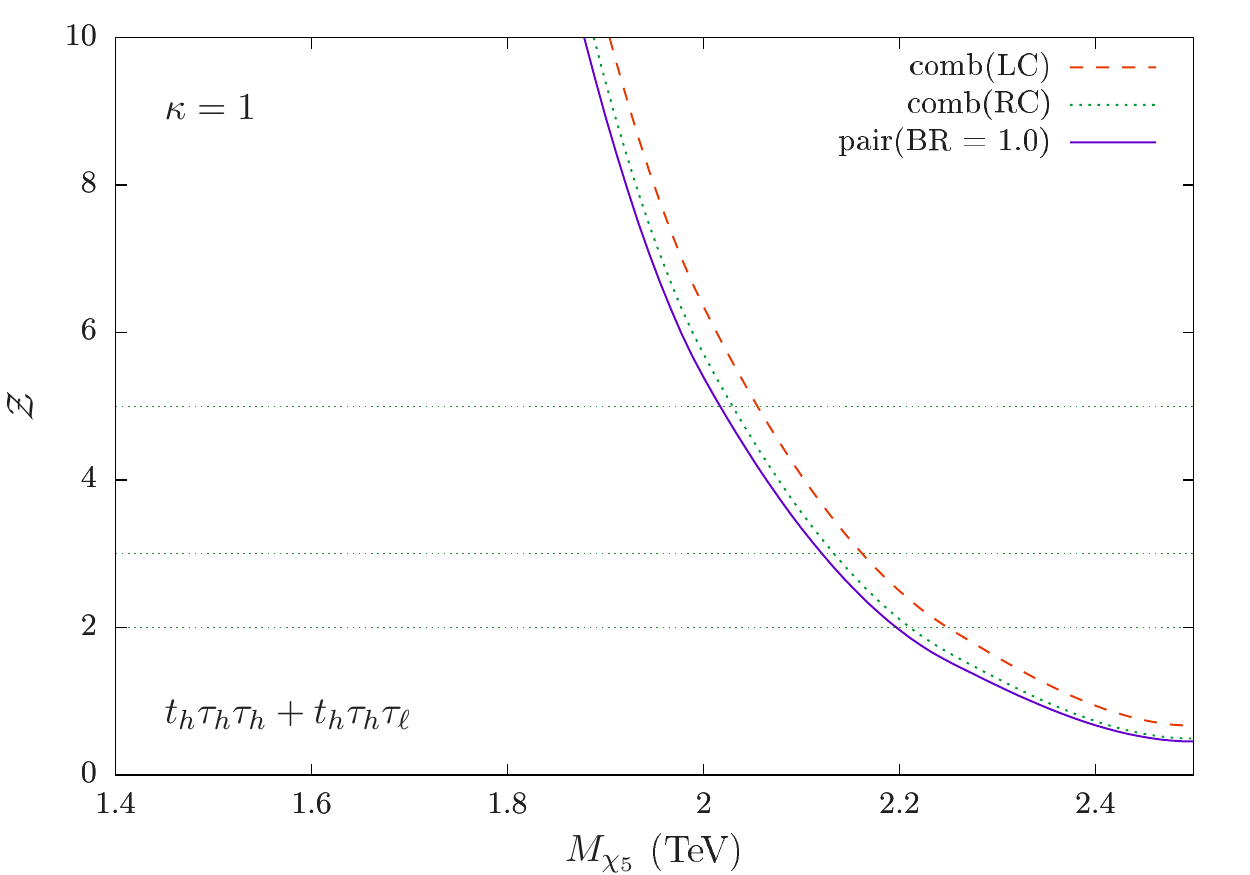}\label{fig:xset104}}\\
\subfloat[\quad\quad\quad(e)]{\includegraphics[width=0.9\columnwidth, height=4.5cm]{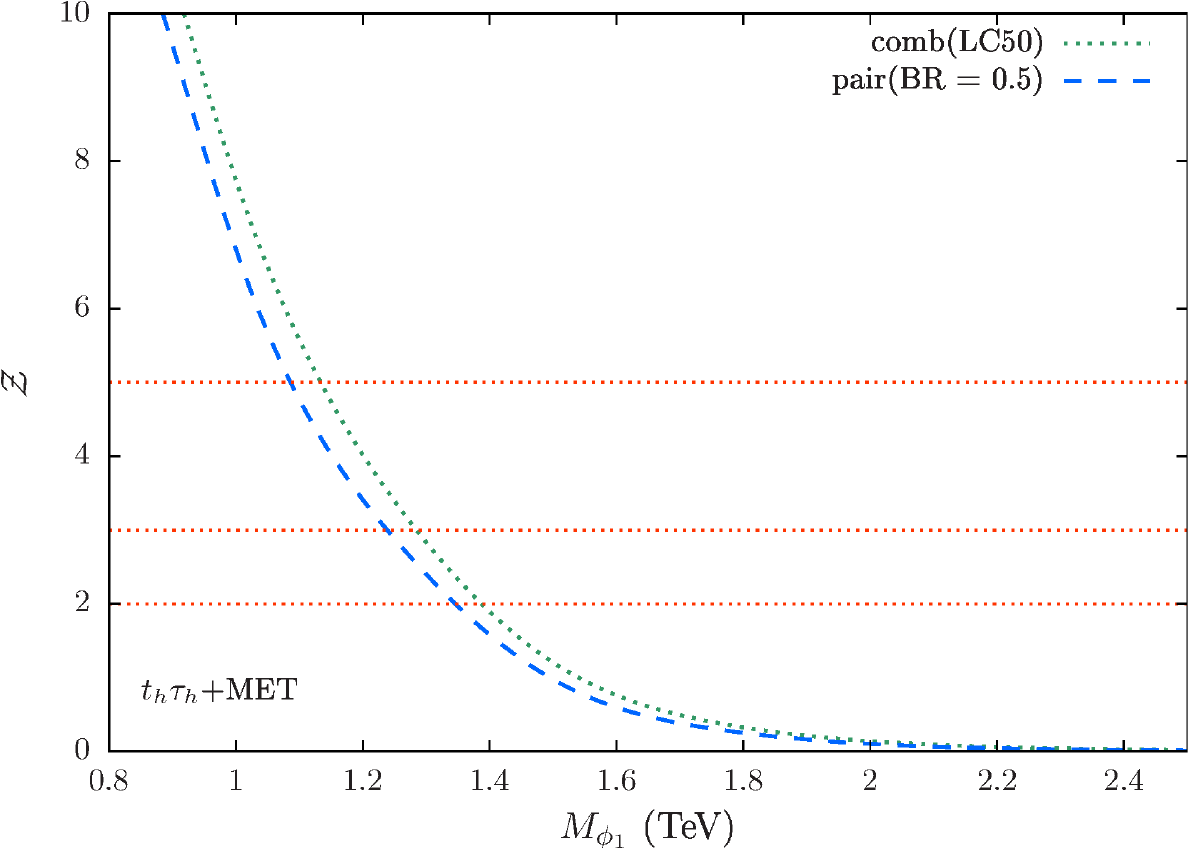}\label{fig:xset201}}\quad\quad\quad\quad
\subfloat[\quad\quad\quad(f)]{\includegraphics[width=0.9\columnwidth, height=4.5cm]{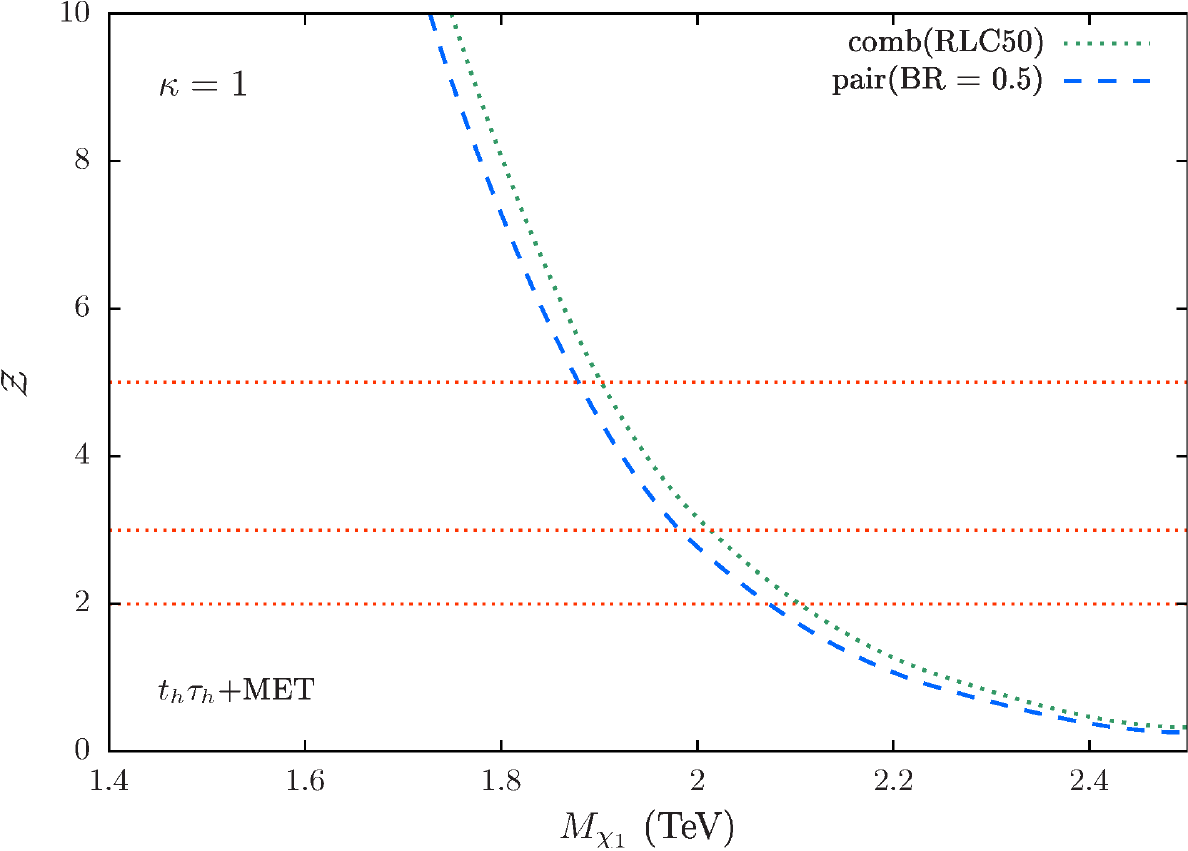}\label{fig:xset202}}\\
\subfloat[\quad\quad\quad(g)]{\includegraphics[width=0.9\columnwidth, height=4.5cm]{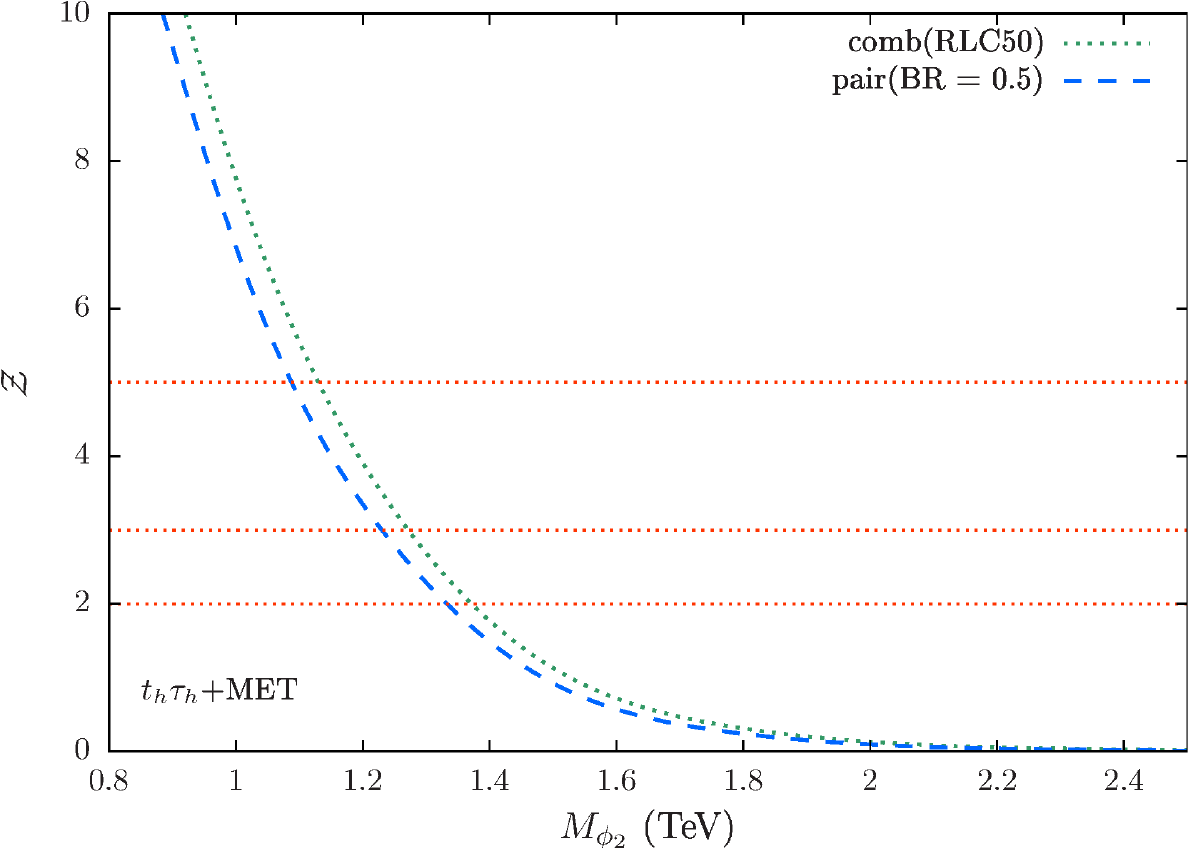}\label{fig:xset203}}\quad\quad\quad\quad
\subfloat[\quad\quad\quad(h)]{\includegraphics[width=0.9\columnwidth, height=4.5cm]{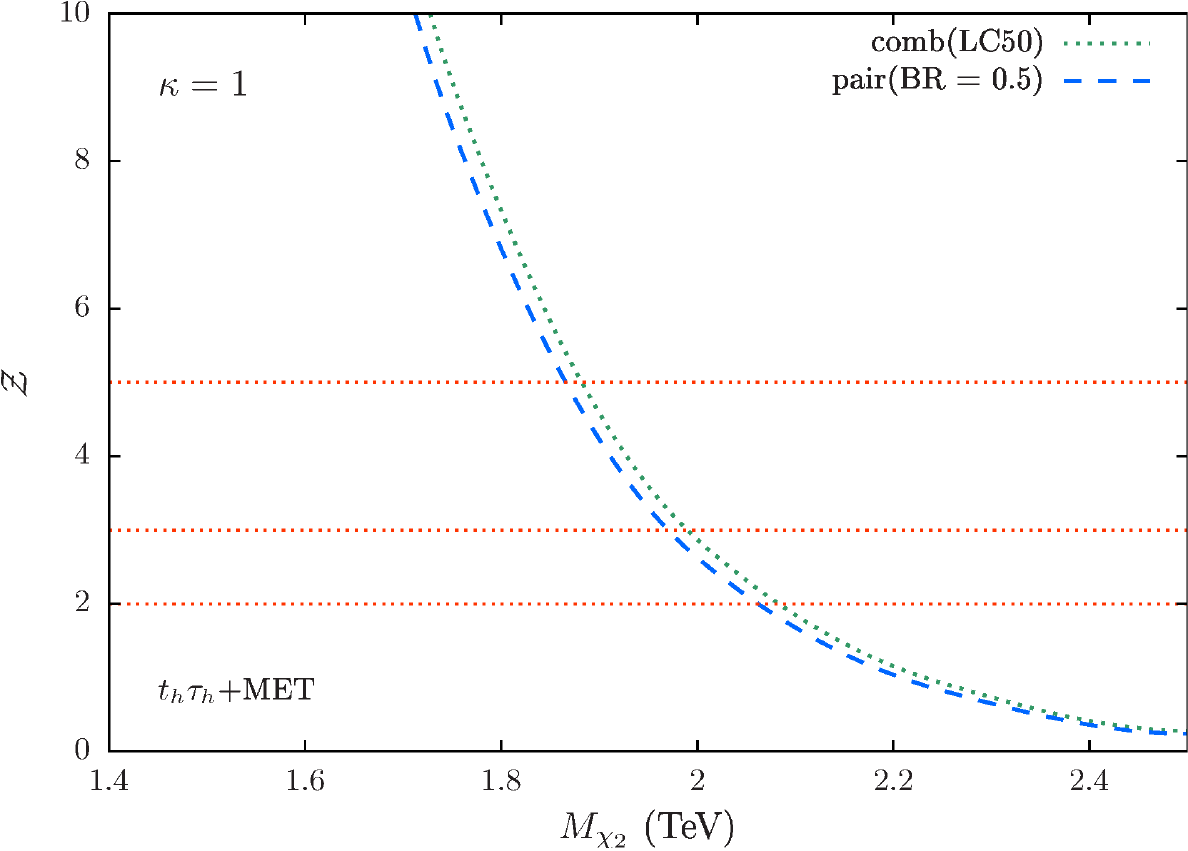}\label{fig:xset204}}
\caption{Expected significance $\mc Z$ for observing the $\phi_{i}$/$\chi_{i}$ signals over the SM backgrounds. 
They are plotted as functions of their masses for $3$ ab$^{-1}$ of integrated luminosity at the 14 TeV HL-LHC for different coupling scenarios. 
The `comb' implies the combined pair and single production processes. We have shown the pair production significance with $\textrm{BR}=50$\% and $\textrm{BR}=100$\%.
We have considered $\lambda$, $\Lm=1$ when computing the signals.}
\label{fig:xset1}
\end{figure*}

\begin{figure*}[]
\centering
\captionsetup[subfigure]{labelformat=empty}
\subfloat[\quad\quad\quad(a)]{\includegraphics[width=0.9\columnwidth, height=4.5cm]{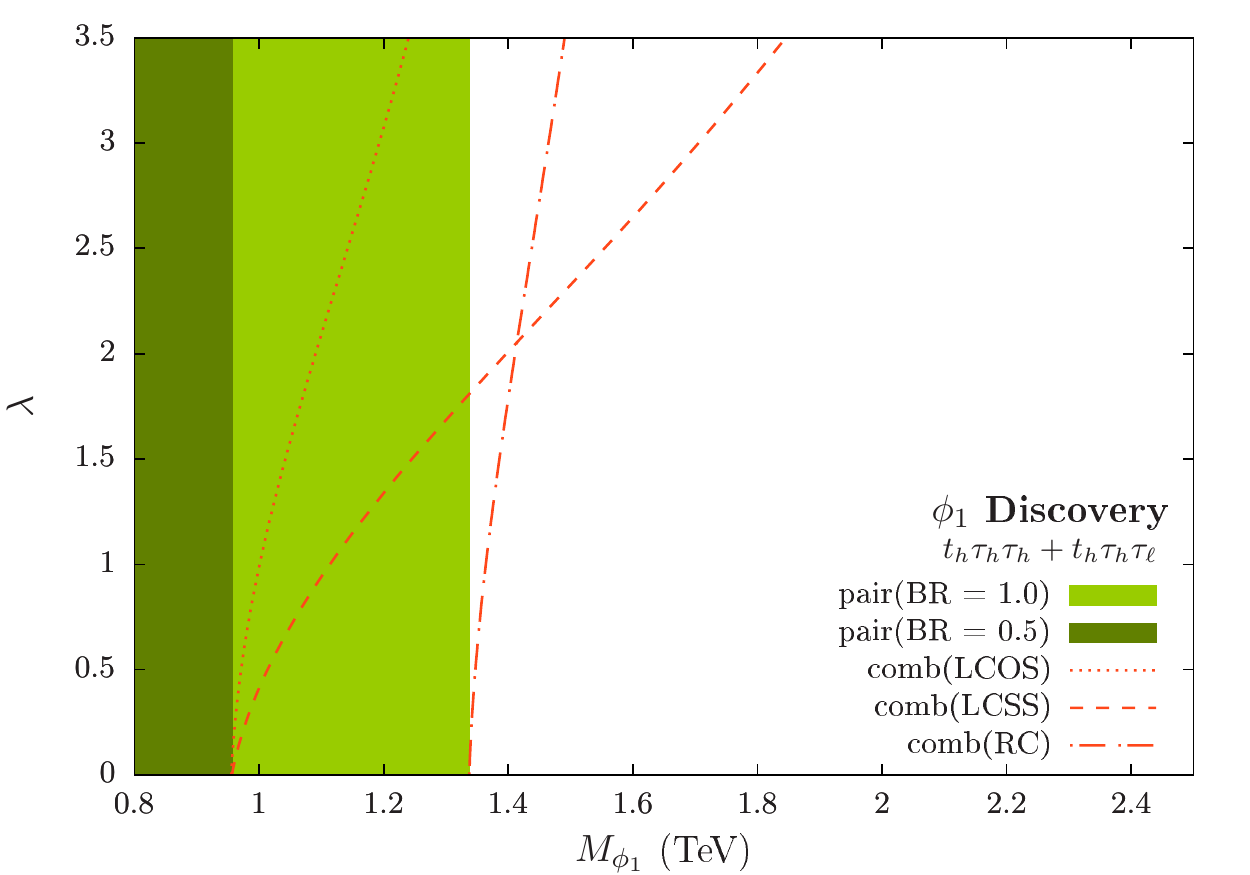}\label{fig:xlams1z5}}\quad\quad\quad\quad
\subfloat[\quad\quad\quad(b)]{\includegraphics[width=0.9\columnwidth, height=4.5cm]{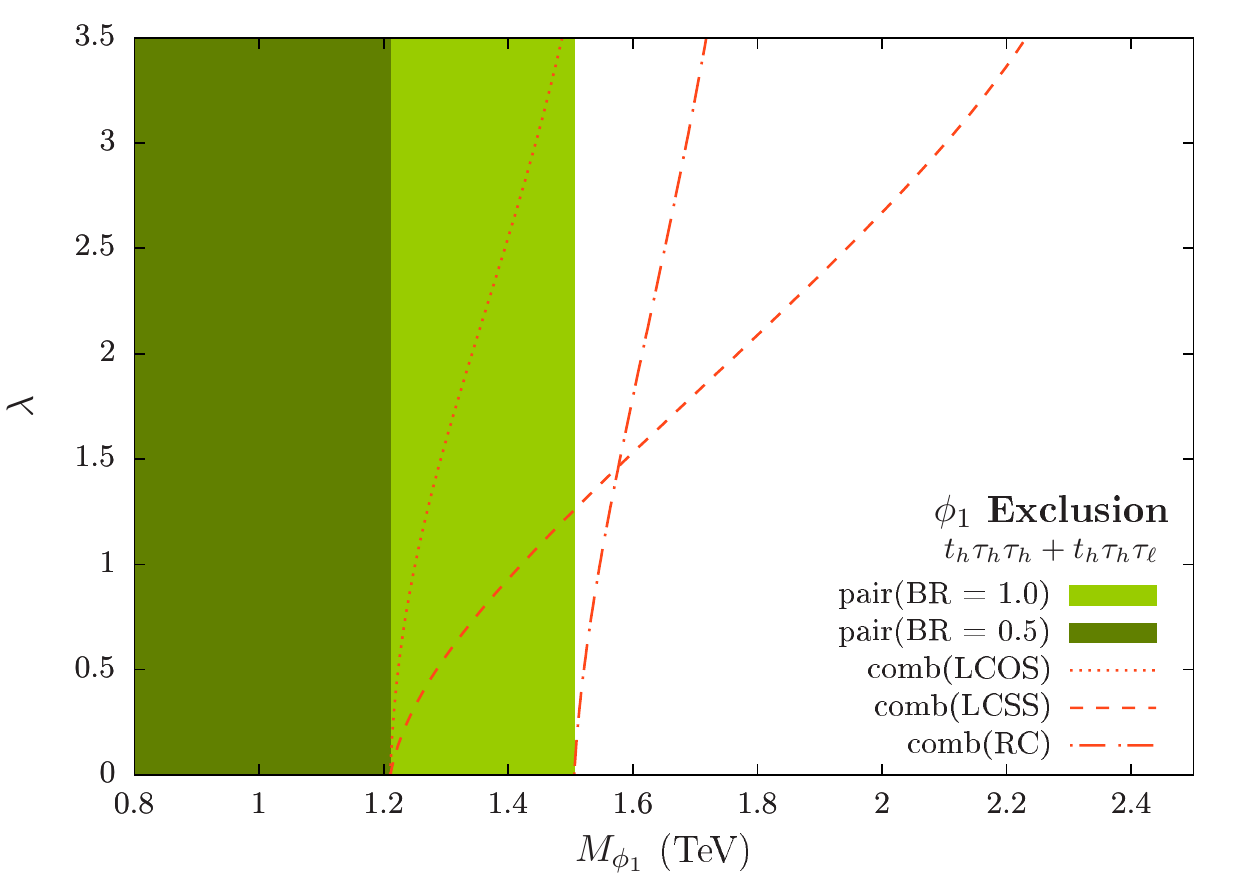}\label{fig:xlams1z2}}\\
\subfloat[\quad\quad\quad(c)]{\includegraphics[width=0.9\columnwidth, height=4.5cm]{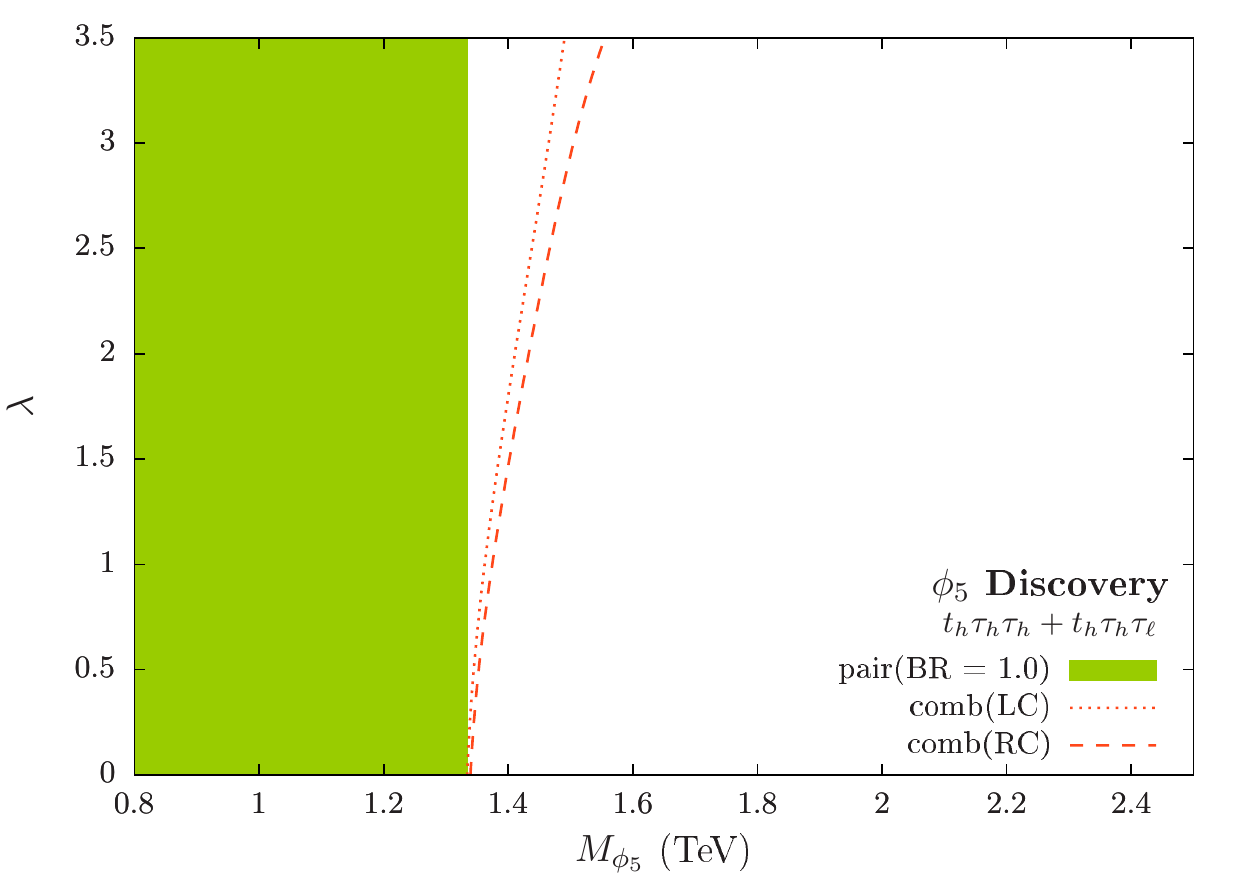}\label{fig:xlams5z5}}\quad\quad\quad\quad
\subfloat[\quad\quad\quad(d)]{\includegraphics[width=0.9\columnwidth, height=4.5cm]{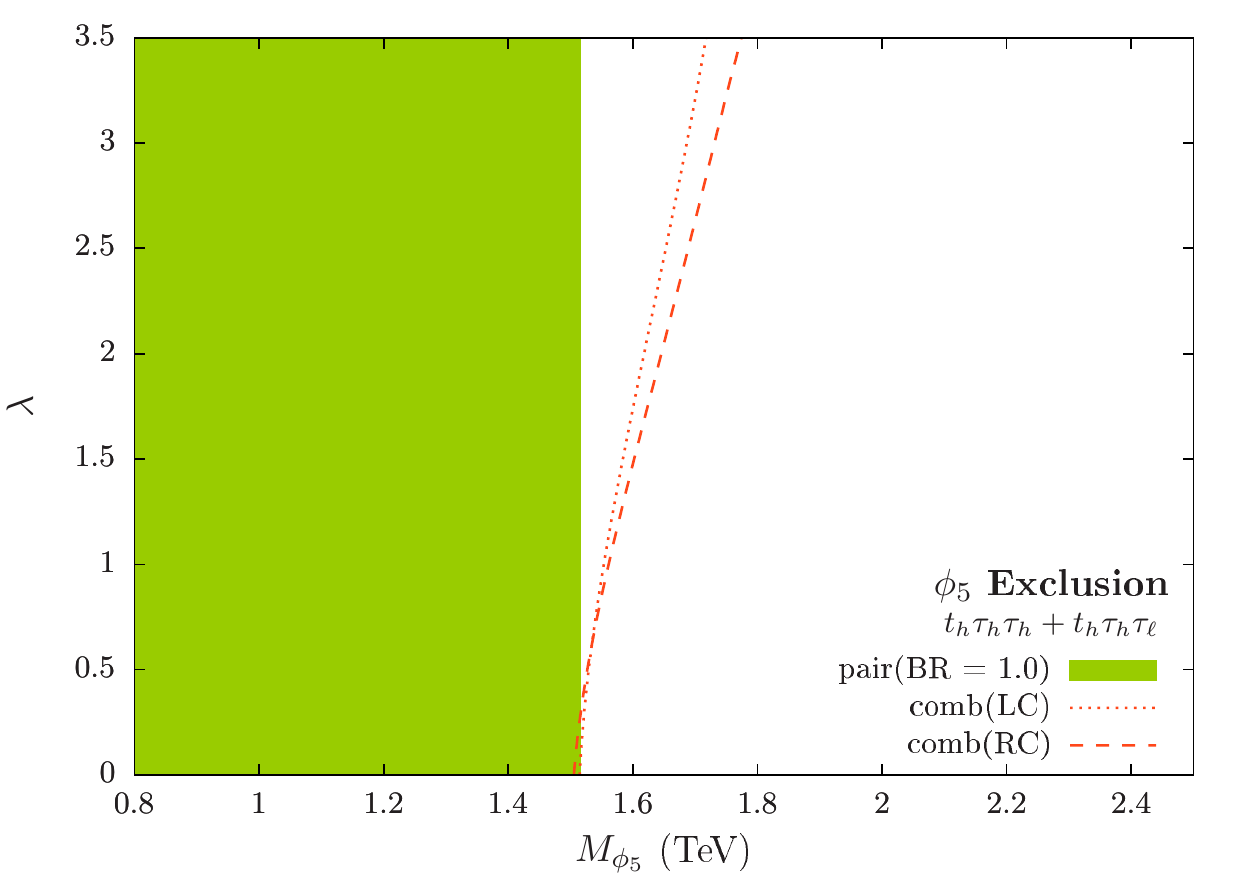}\label{fig:xlams5z2}}\\
\subfloat[\quad\quad\quad(e)]{\includegraphics[width=0.9\columnwidth, height=4.5cm]{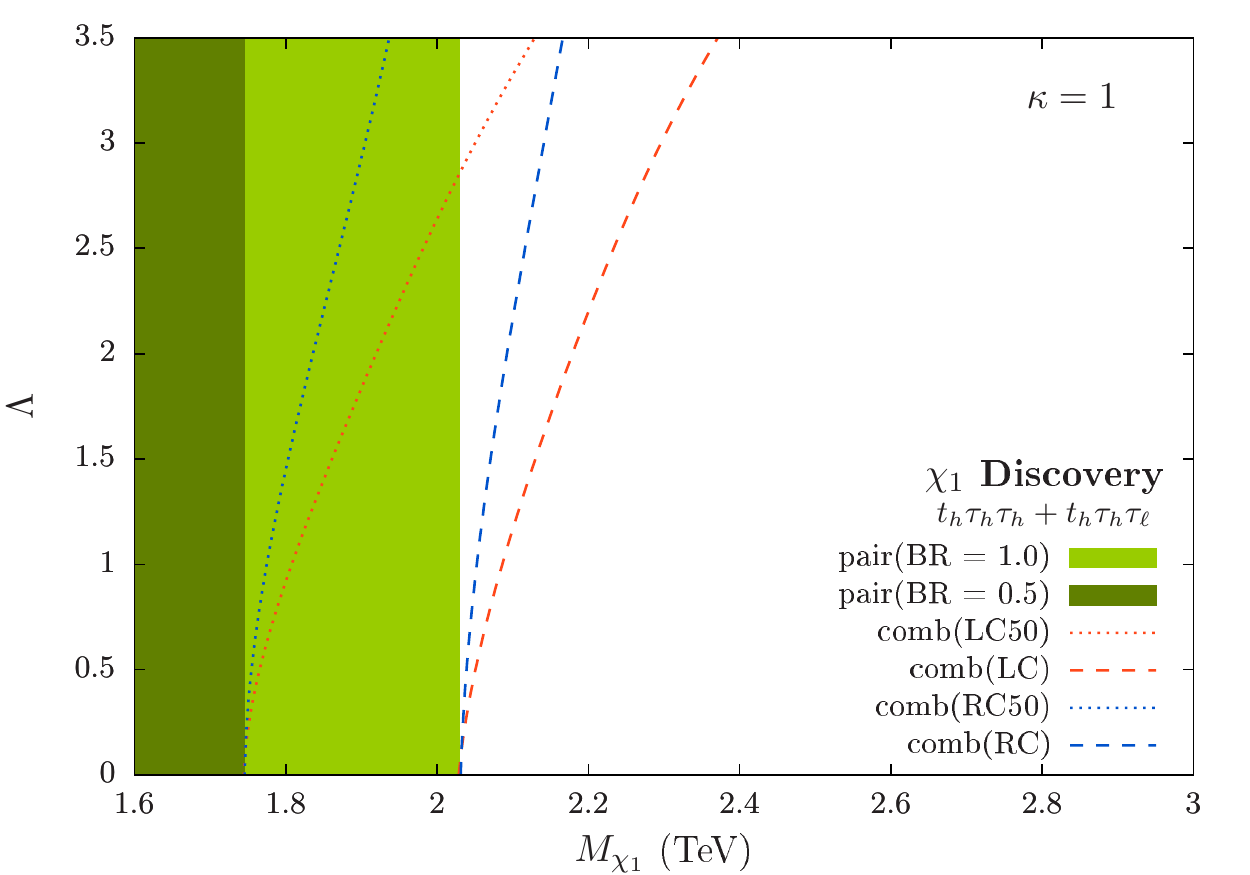}\label{fig:xlamv1k1z5}}\quad\quad\quad\quad
\subfloat[\quad\quad\quad(f)]{\includegraphics[width=0.9\columnwidth, height=4.5cm]{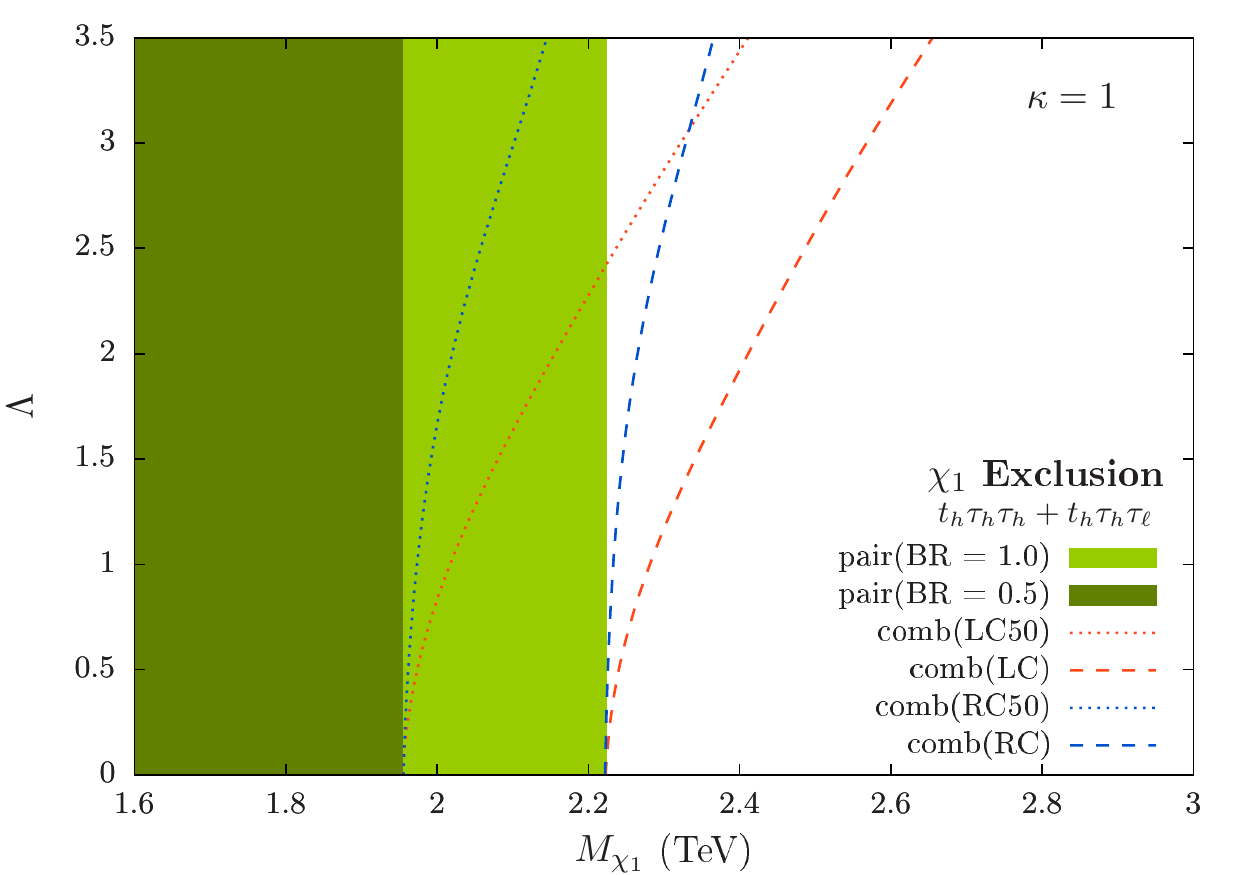}\label{fig:xlamv1k1z2}}\\
\subfloat[\quad\quad\quad(g)]{\includegraphics[width=0.9\columnwidth, height=4.5cm]{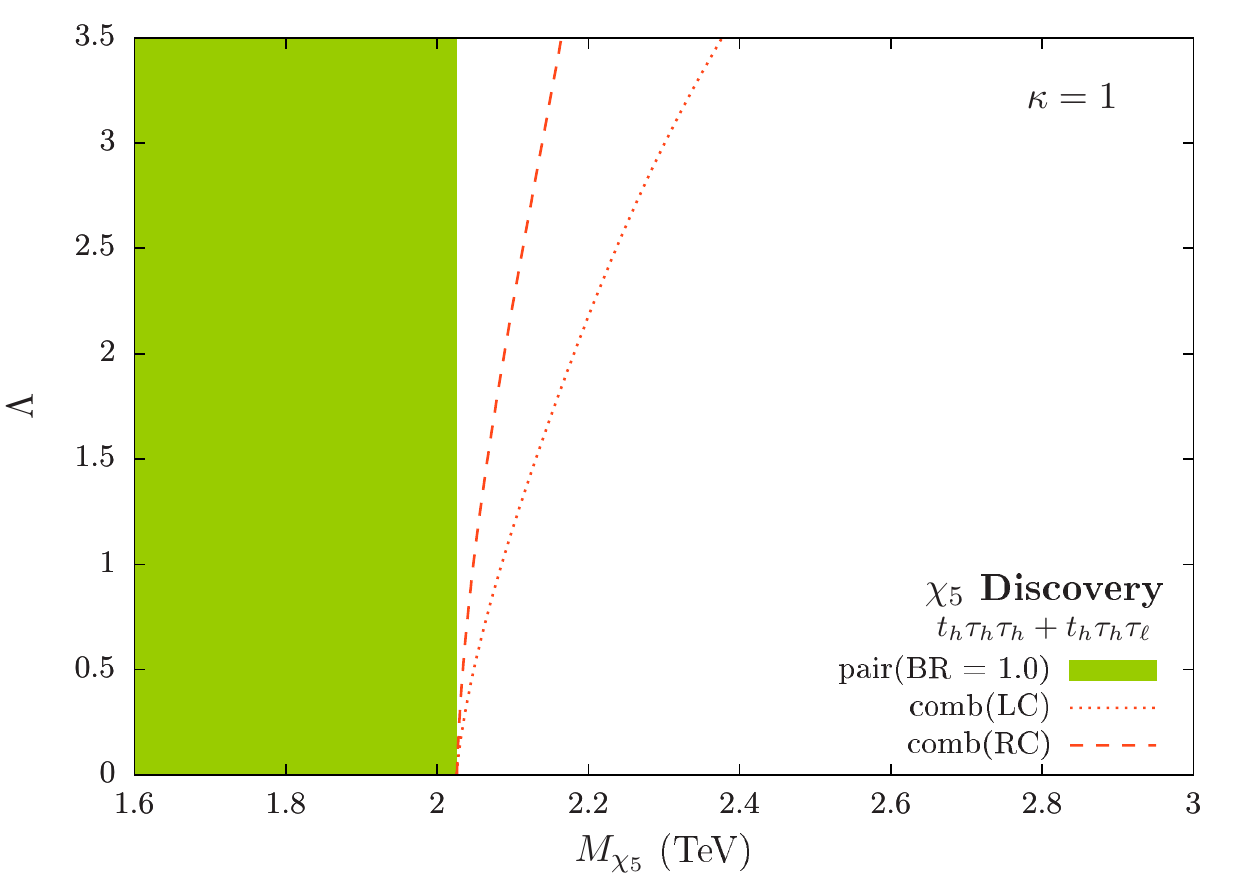}\label{fig:xlamv5k1z5}}\quad\quad\quad\quad
\subfloat[\quad\quad\quad(h)]{\includegraphics[width=0.9\columnwidth, height=4.5cm]{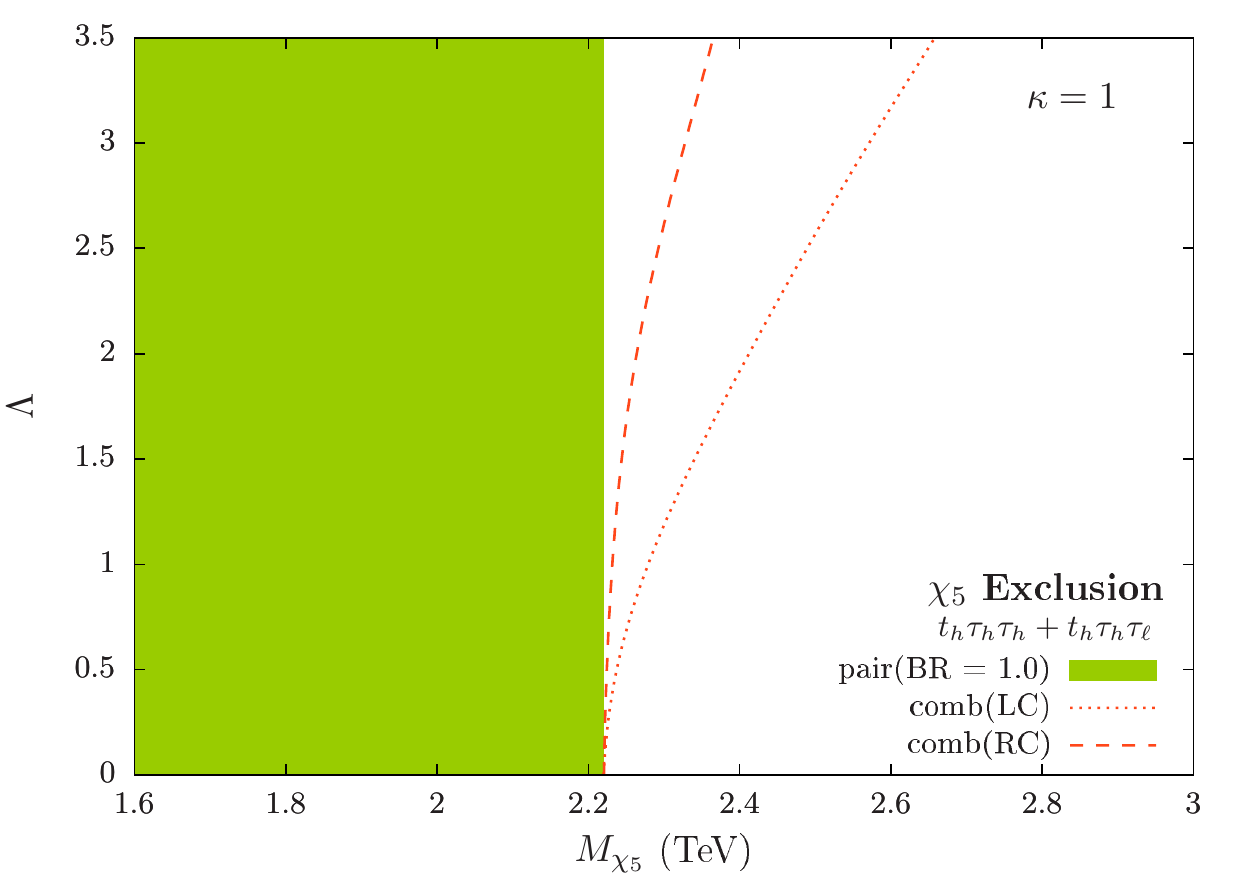}\label{fig:xlamv5k1z2}}
\caption{The $5\sg$ ($2\sg$) discovery (exclusion) reaches in the mass-coupling plane. 
These plots describe the lowest values of couplings needed to observe LQ signals with $5\sg$ and $2\sg$ significance as functions of masses with $3$ ab$^{-1}$ of integrated luminosity. The pair-production-only regions for $50\%$ and $100\%$ BRs in the $\chi/\phi\to t\tau$ decay mode are shown in green. The pair production processes are insensitive to couplings. 
}
\label{fig:xlamv1s1z5r}
\end{figure*}

\begin{figure*}[]
\centering
\captionsetup[subfigure]{labelformat=empty}
\subfloat[\quad\quad\quad(a)]{\includegraphics[width=0.9\columnwidth, height=4.5cm]{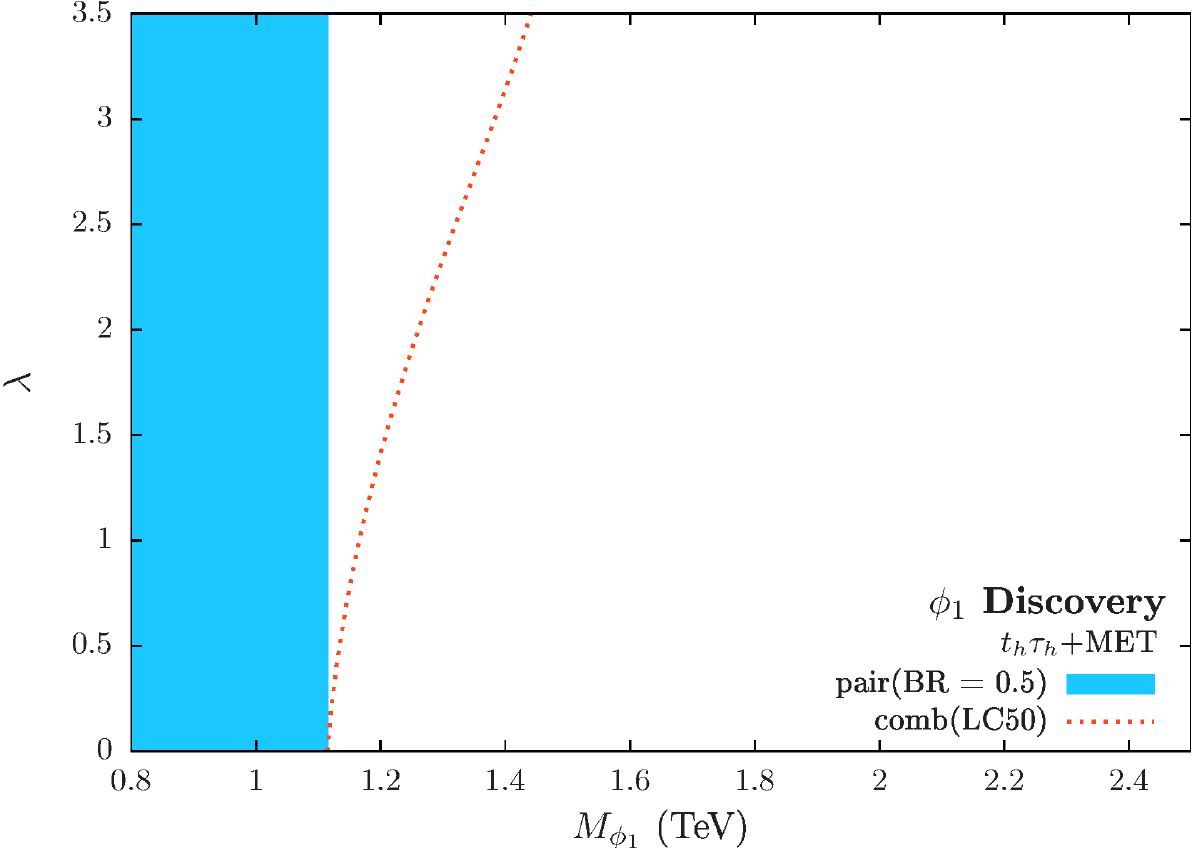}\label{fig:xlam2s1z5}}\quad\quad\quad\quad
\subfloat[\quad\quad\quad(b)]{\includegraphics[width=0.9\columnwidth, height=4.5cm]{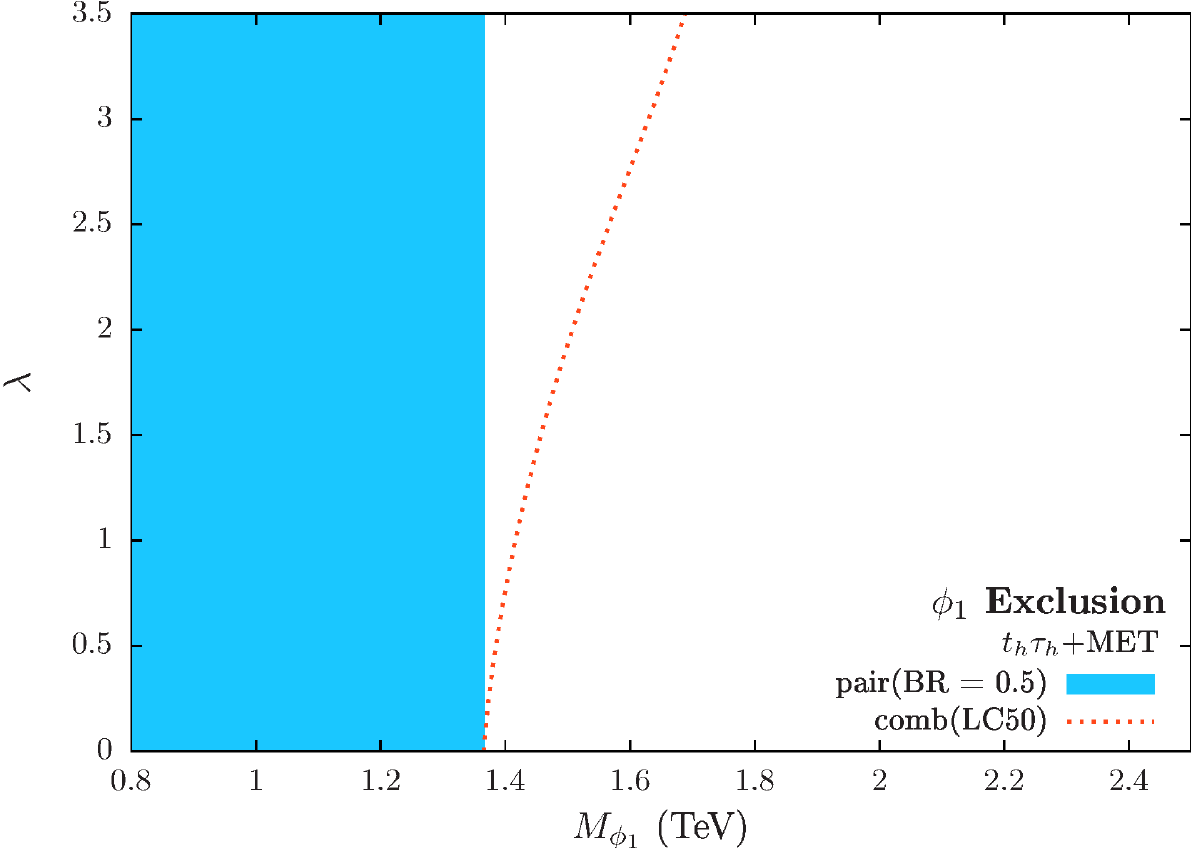}\label{fig:xlam2s1z2}}\\
\subfloat[\quad\quad\quad(c)]{\includegraphics[width=0.9\columnwidth, height=4.5cm]{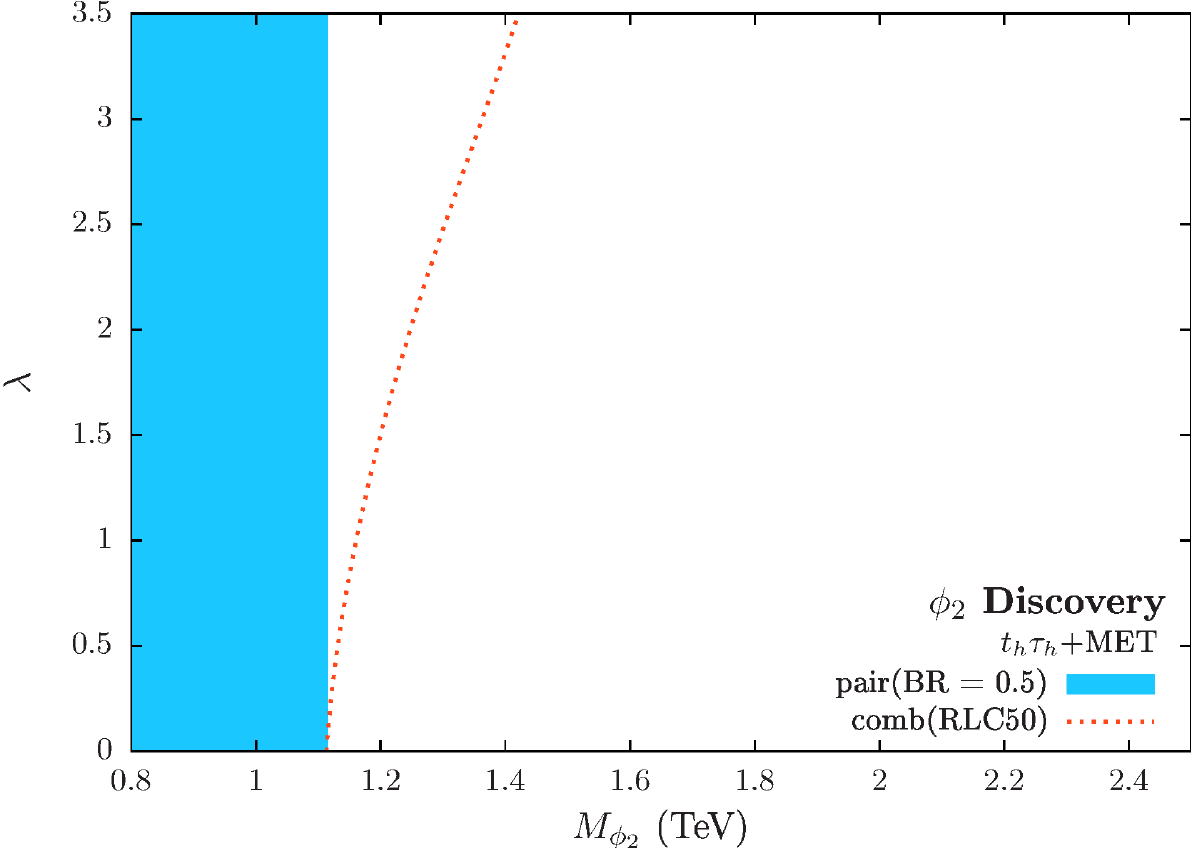}\label{fig:xlam2s2z5}}\quad\quad\quad\quad
\subfloat[\quad\quad\quad(d)]{\includegraphics[width=0.9\columnwidth, height=4.5cm]{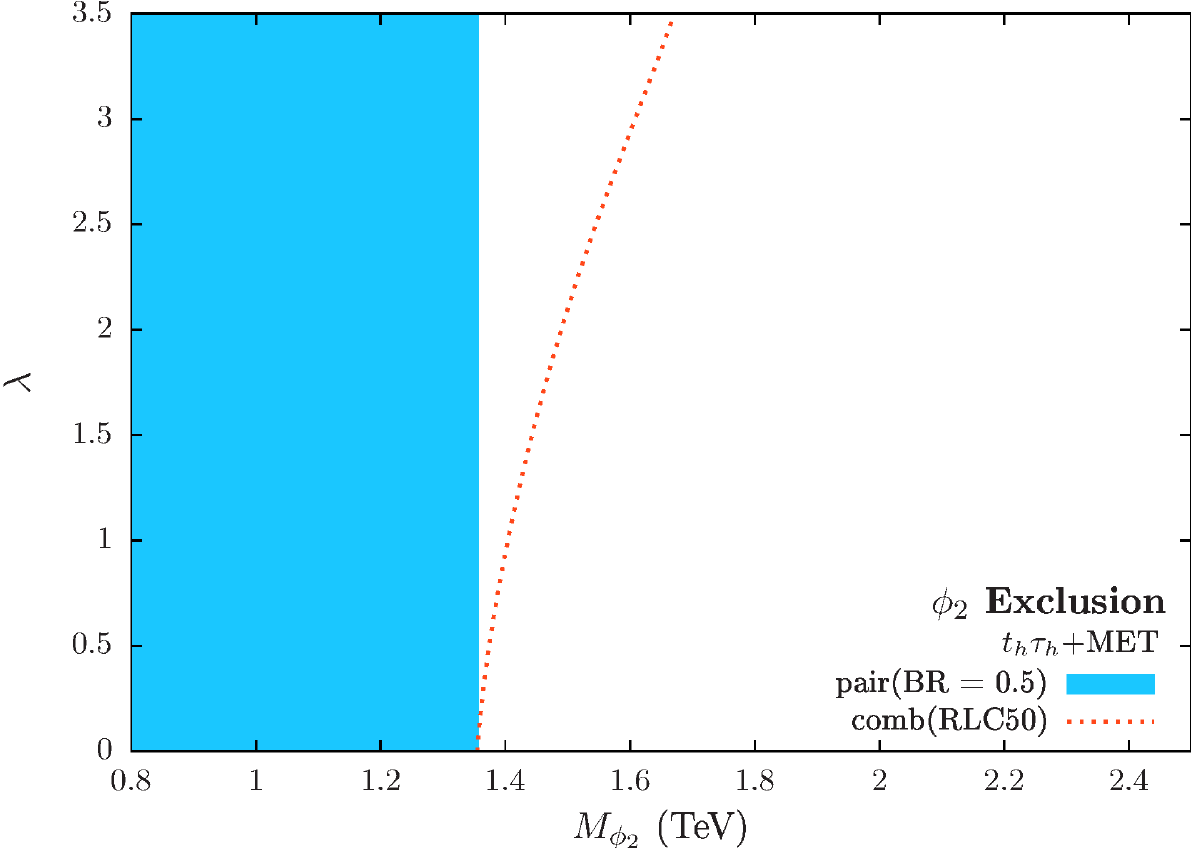}\label{fig:xlam2s2z2}}\\
\subfloat[\quad\quad\quad(e)]{\includegraphics[width=0.9\columnwidth, height=4.5cm]{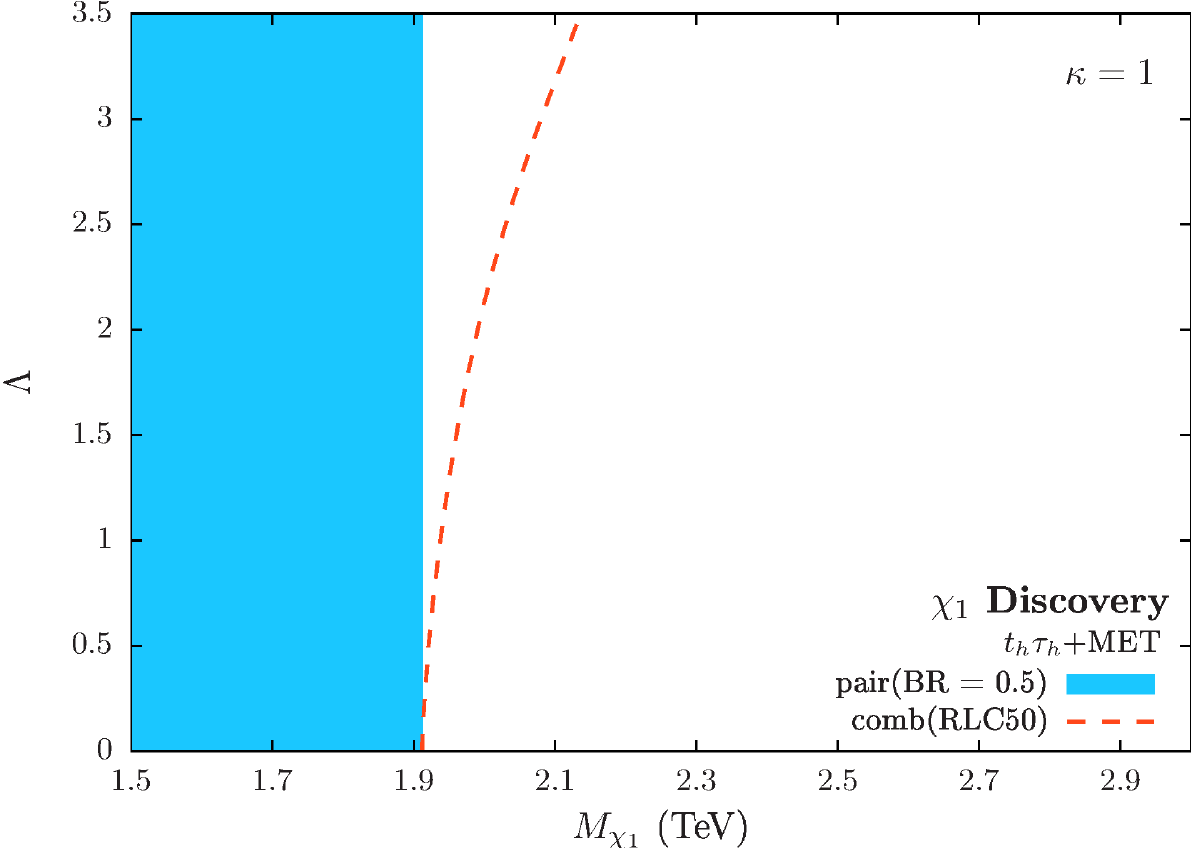}\label{fig:xlam2v1k1z5}}\quad\quad\quad\quad
\subfloat[\quad\quad\quad(f)]{\includegraphics[width=0.9\columnwidth, height=4.5cm]{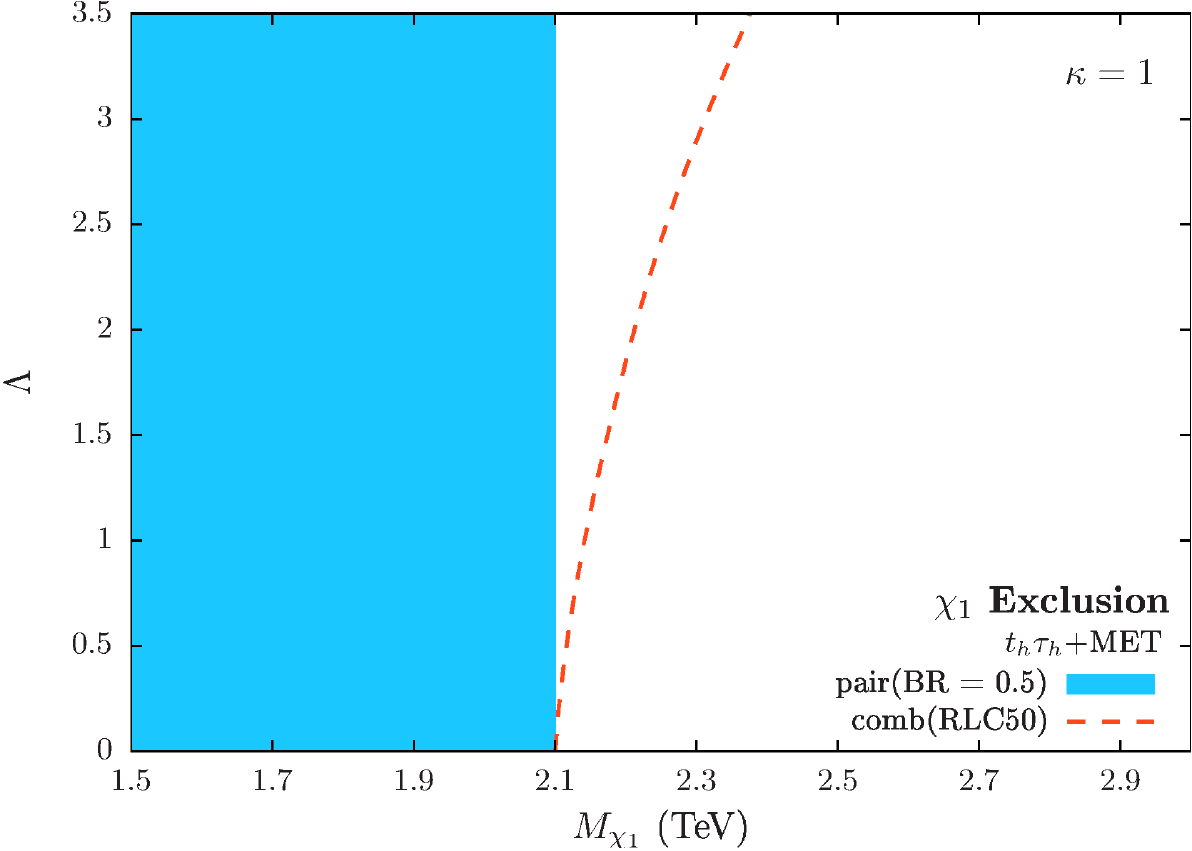}\label{fig:xlam2v1k1z2}}\\
\subfloat[\quad\quad\quad(g)]{\includegraphics[width=0.9\columnwidth, height=4.5cm]{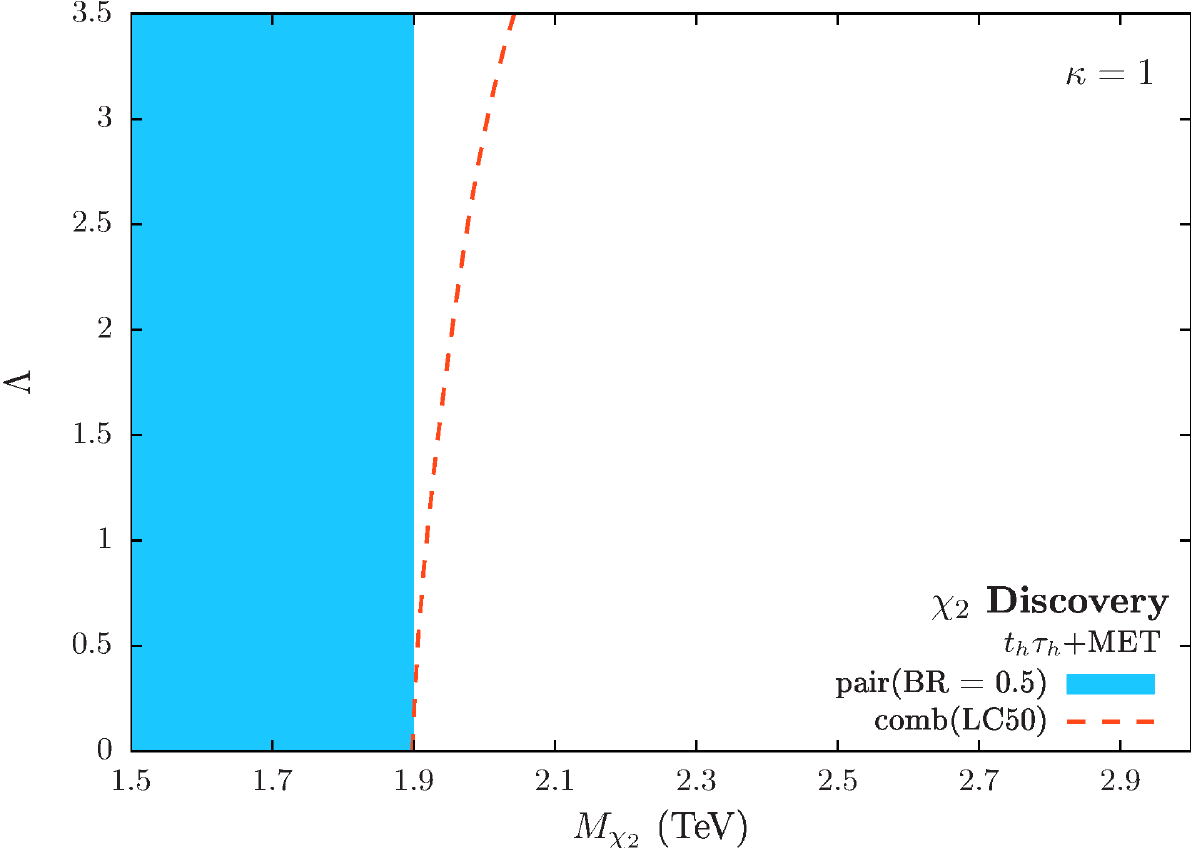}\label{fig:xlam2v2k1z5}}\quad\quad\quad\quad
\subfloat[\quad\quad\quad(h)]{\includegraphics[width=0.9\columnwidth, height=4.5cm]{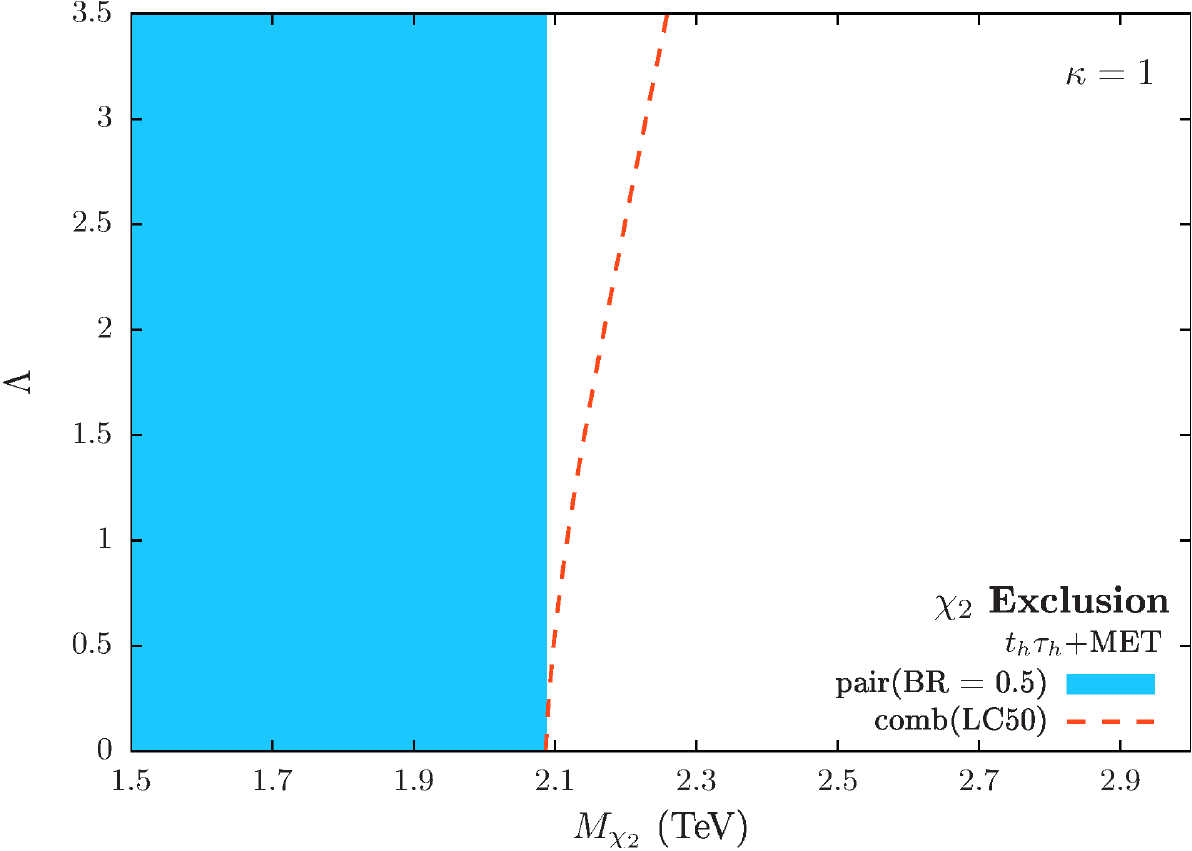}\label{fig:xlam2v2k1z2}}
\caption{
Same as Fig.~\ref{fig:xlamv1s1z5r} for the $t_h\tau_h+\textrm{MET}$ channel.
The pair-production-only regions for $50\%$ BRs in the $\chi/\phi\to t\tau$ and $b\nu$ decay mode are shown in blue. 
}
\label{fig:xlamtbTauNu5S}
\end{figure*}

\noindent
We use the following definition of statistical significance $\mc{Z}$
\begin{align}
\mc{Z} = \sqrt{2\lt(N_S+N_B\rt)\ln\lt(\frac{N_S+N_B}{N_B}\rt)-2N_S}\, ,\label{eq:sig}
\end{align}
where the number of signal and background events surviving the final selection cuts (as listed in Table~\ref{tab:cuts}) are denoted by $N_S$ and $N_B$, respectively. 
We show the expected significance as functions of LQ masses in Fig.~\ref{fig:xset1}. Figs.~\ref{fig:xset101}--\ref{fig:xset104} are for the $t_h\tau_h\tau_h+t_h\tau_h\tau_\ell$ channel and Figs.~\ref{fig:xset201}--\ref{fig:xset204}
are for the $t_h\tau_h+\textrm{MET}$ channel.
These plots are obtained for the $14$ TeV LHC with $3$~ab$^{-1}$ of integrated luminosity.
We also display the masses corresponding to $\mc{Z} = 2$, $3$, and $5$ in Tables~\ref{tab:sig2}, \ref{tab:sig1}, and \ref{tab:sig3}.
As discussed earlier, the choice of $\kp$ affects the pair and some single productions of vLQs and hence can significantly change the reach.
In the figures we only present the results for $\kp=1$ but in Tables~\ref{tab:sig1} and \ref{tab:sig3}, we show the mass limits for another choice of, namely,  $\kp=0$.
We have used $\lm,\Lm = 1$ to estimate the significance for the combined signal (i.e., the pair and single production events together). 

The figures are largely self-explanatory, and so, we only discuss some important points here. In Fig.~\ref{fig:xset101}, we show the expected significance $\mc{Z}$ in the $t_h\tau_h\tau_h+t_h\tau_h\tau_\ell$ channel as functions of $M_{\phi_1}$ (i.e., the mass of the charge-$1/3$ sLQ) in different scenarios. 
Notice that one can probe  higher $M_{\phi_1}$ values in the combined LCSS scenario than the combined LCOS scenario even though the decays $\phi_1\to t\tau$ and $\phi_1\to b\nu$ share $50$\% BR each in both scenarios (see Table~\ref{tab:benchmark}). 
The difference appears from the (constructive/destructive) interference in single production diagrams in these two scenarios that we discussed earlier. The destructive interference in the LCOS scenario takes the combined curve closer to the `pair (BR $=1$)' curve. In the RC scenario, $\phi_1$ has $100$\% BR in the $\phi_1\to t\tau$ decay mode. However, the single production cross section is small in this scenario (see Fig.~\ref{fig:xsec01}). As a result, the combined reach is only marginally improved than the pair production only case with $\textrm{BR}=1$. 

For the charge-$1/3$ vLQ $\chi_1$, there are no LCOS and LCSS scenarios. In Fig.~\ref{fig:xset102}, the LC50 and RC50 curves represent the cases where the BR of $\chi_{1}\to t\tau$ mode is $50$\%. Such a scenario is possible if there are other decay modes of $\chi_1$ that play no role in our analysis beyond modifying the BR. Hence, we show these plots to estimate how the significance would vary with the BR. 
The signal significance for $\phi_5$ ($\chi_5$) is similar to that of $\phi_1$ ($\chi_1$) with $100$\% BR to $t\tau$ mode. 

In the CMS analysis~\cite{Sirunyan:2020zbk}, the $tb\tau\nu+t\tau\nu$ channel is considered for the charge-$2/3$ vLQ ($\chi_2$) or the charge-$1/3$ sLQ ($\phi_1$). However, the same final state can also arise from a charge-$2/3$ sLQ ($\phi_2$) or a charge-$1/3$ vLQ ($\chi_1$) as well. We obtain the HL-LHC reach in the $t_h\tau_h+\textrm{MET}$ channel for both charge-$1/3$ and $2/3$ sLQs and vLQs ($\phi_{1,2}$ and  $\chi_{1,2}$). In this channel, the interference effect for $\phi_1$ is not visible, and therefore, we club the LCOS and LCSS scenarios as LC50 in~Fig.~\ref{fig:xset201}). The LC50 significance curve lies somewhere between the LCSS and LCOS curves in the $2\ta$ channel. For $\chi_1$, the RLCOS and RLCSS (clubbed as RLC50) lead to similar significance as BR $=50$\% cases in the $2\ta$ channel. The reach for $\phi_2$ in the RLC50 scenario  and for $\chi_2$ the LC50 scenario  are shown in Figs.~\ref{fig:xset203} and \ref{fig:xset204}, respectively.

We can parametrise the coupling dependence of combined signal cross section (in any channel) as follows:
\be
\sg_{\rm combined} \approx \sg_{\rm pair}(M_\phi) + \lm^2 \sg_{\rm single}(\lm=1,M_\phi).
\ee
The above equation also holds for vLQs with a fixed $\kp$.
Using those relations, we can obtain the HL-LHC $5\sg$ discovery reach and $2\sg$ exclusion limits in the
coupling-mass planes. We present these plots in Figs.~\ref{fig:xlamv1s1z5r} 
and \ref{fig:xlamtbTauNu5S} for the $1\tau$ and $2\tau$ signatures. 
These plots show the lowest values of LQ-$q$-$\ell$ couplings needed to observe the LQ signatures as functions of LQ masses with $5\sg$ confidence level for discovery. For the exclusion plots, all points above the
curves can be excluded with $95\%$ confidence level at the HL-LHC. These plots are significant from the perspective of the $B$-meson anomalies. For example, the $\lm$ in the LCOS curves in Figs.~\ref{fig:xlams1z5} and~\ref{fig:xlams1z2} represent the $y_{1\ 33}^{LL}$ coupling of $S_1$ or the $\Lm$ in the LC50 curves in Figs.~\ref{fig:xlam2v2k1z5} and~\ref{fig:xlam2v2k1z2} 
is the $x_{1\ 33}^{LL}$ coupling of $U_1$. Hence, these plots show how far the LHC can probe the couplings required to explain the anomalies.

In our analysis so far, we have ignored the possible systematic errors. In practice, experiments have to account for them. However, as we argue in Appendix~\ref{sec:appendixA}, our results can be taken as a reasonable conservative estimate of the projected experimental limits. Because of more precise (e.g., data-driven) estimation of the background processes  and the use of profile likelihood ratios to estimate the CLs instead of approximate formulae of signal significance, we expect the actual experimental limits to be better than our estimations. The use of advanced analysis techniques (like machine learning) can improve the results further.

\section{Summary and conclusions}\label{sec:End}

\noindent
In this paper, we have considered all scalar and vector LQs that can decay to $t\tau$ (applicable for charge-$1/3$ and charge-$5/3$ LQs) or $t\nu$ pair (applicable for charge-$2/3$ LQs). Our choice of the decay modes involving a top quark is motivated from the fact that a top quark from a heavy LQ decay would be boosted. The features of the boosted top can then be used to search for the LQs. In our previous papers~\cite{Chandak:2019iwj,Bhaskar:2020gkk}, we obtained the HL-LHC prospects for LQs that can decay to a boosted top quark and a light lepton (i.e. $e$ or $\mu$). Here, we treat the case of the $\tau$-lepton separately since the hadronic and leptonic decays of $\tau$-leptons complicate the analysis.
Therefore, the search strategies adopted here are completely different than those applied for the light leptons. Moreover, in this paper, we extend our study by including an asymmetric channel for charge-$1/3$ and charge-$2/3$ LQs.

We considered two different signatures: (a) at least one hadronically decaying boosted top quark and two high-$p_T$ tau leptons. This signature is denoted as $t_h\tau_h\tau_h+t_h\tau_h\tau_\ell$, according to the $\tau$ decay modes involved, and (b) at least one hadronically decaying boosted top quark and one hadronically decaying high-$p_T$ tau lepton with some missing energy ($t_h\tau_h+\textrm{MET}$). In both signatures, we used both standard jets (AK4-jets) and fatjets (AK8-fatjets) in the event selection criteria. For example, presence of $\tau_h$(s) and a $b$-jet in the signals are ensured by tagging the AK4-jets, and the AK8-fatjets are used to form the hadronic top quark. If an AK4-jet is tagged as $\tau_h$, we make sure that it does not come from the AK8-fatjet identified as $t_h$. But the $b$-tagged jet may or may not be a part of $t_h$. The presence of $b$-jet helps us to tame the $V+$ jets backgrounds.

An important point in our analysis is that the two signatures are inclusive in nature. Hence, they can keep both pair and single production events with substantial fractions, i.e., our selection criteria are not optimized for pair or single productions separately, rather they are optimized for their combination. Such a combined-signal strategy can significantly improve the exclusion limits or discovery reach for the LQs. Because of the messy nature of final states due to the presence of $\tau$, the reach for the third-generation LQs is lower than was obtained in Refs.~\cite{Chandak:2019iwj,Bhaskar:2020gkk} with the light leptons. However, the overall HL-LHC reach goes beyond $1.5$ TeV for sLQs and $2$ TeV for vLQs. We have also mimicked the CMS selection criteria for the boosted or ``fully-merged'' top quark category from Ref.~\cite{Sirunyan:2020zbk}  to compare their analysis with ours. With our analysis, the future LHC prospects for the combined signal stand slightly better than those obtained with the CMS cuts. Therefore, our proposed analysis can act as a good search strategy for the specific subset of LQs we considered.  

The single production processes that enter in the combined signal are three-body single productions whereas the commonly considered ones are two-body single productions. 
We found an interesting interference between single production diagrams for the charge-$1/3$ sLQ in the $pp\to \phi_1\tau j$ channel. We introduced two scenarios (LCSS and LCOS) that can be realized in the singlet ($S_1$) and triplet ($S_3$) sLQ models to demonstrate this effect. 
In the LCSS scenario, due to constructive interference, the discovery reach is appreciably higher than that in the LCOS scenario where the interference is destructive in nature. 
This interference effect is not observed in other LQs.

\acknowledgments 
\noindent 
A.B. and S.M. acknowledge financial support from the Science and
Engineering Research Board (SERB), India under Grant No. ECR/2017/000517.
T.M. is supported by the SERB, India under research Grant No. 
CRG/2018/004889 and an intramural grant from IISER-TVM.


\begin{figure*}[]
\centering
\captionsetup[subfigure]{labelformat=empty}
\subfloat[\quad\quad\quad(a)]{\includegraphics[width=0.9\columnwidth, height=4.5cm]{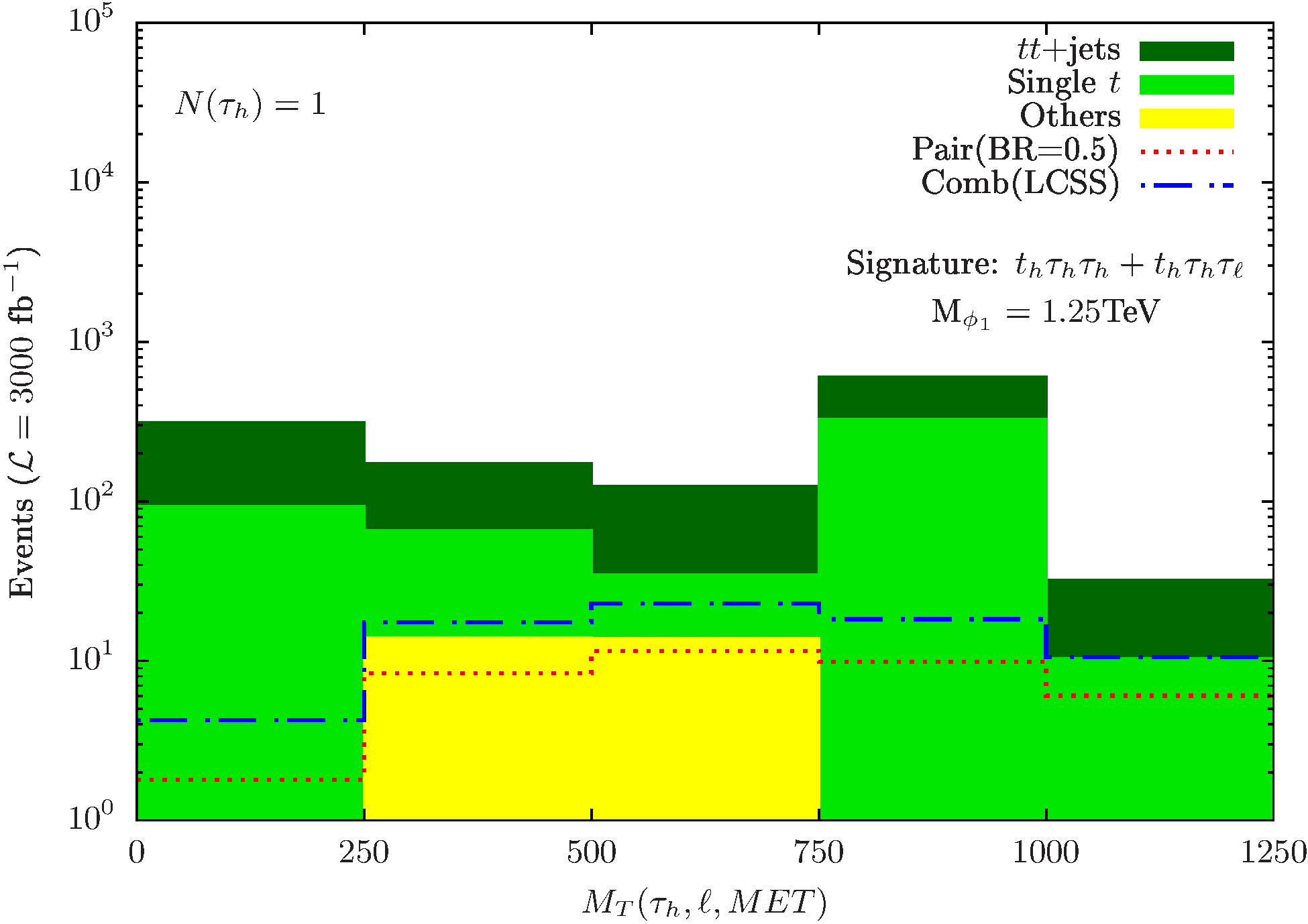}\label{fig:sig1_ntau1}}\quad\quad\quad\quad
\subfloat[\quad\quad\quad(b)]{\includegraphics[width=0.9\columnwidth, height=4.5cm]{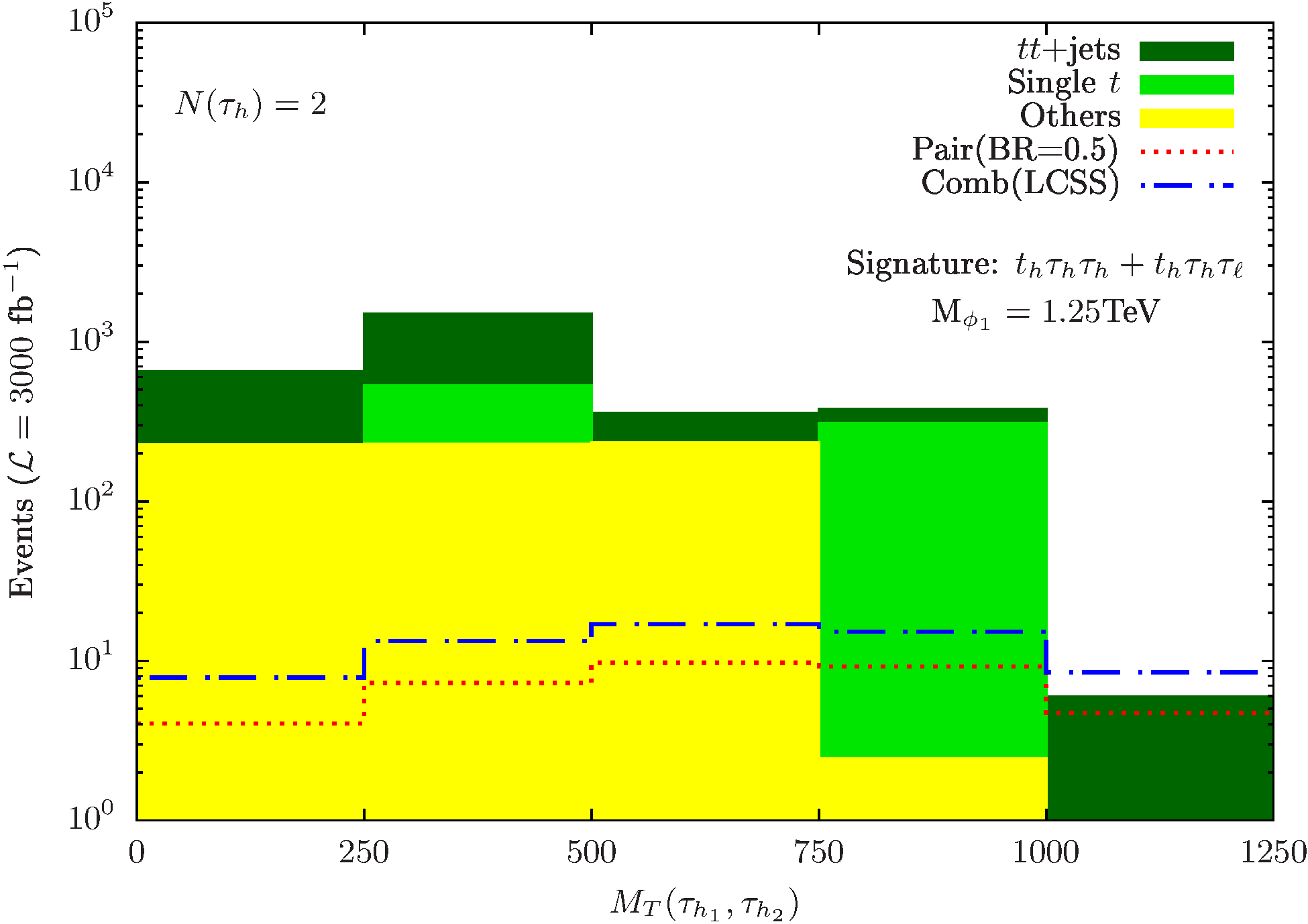}\label{fig:sig1_ntau2}}\\
\subfloat[\quad\quad\quad(c)]{\includegraphics[width=0.9\columnwidth, height=4.5cm]{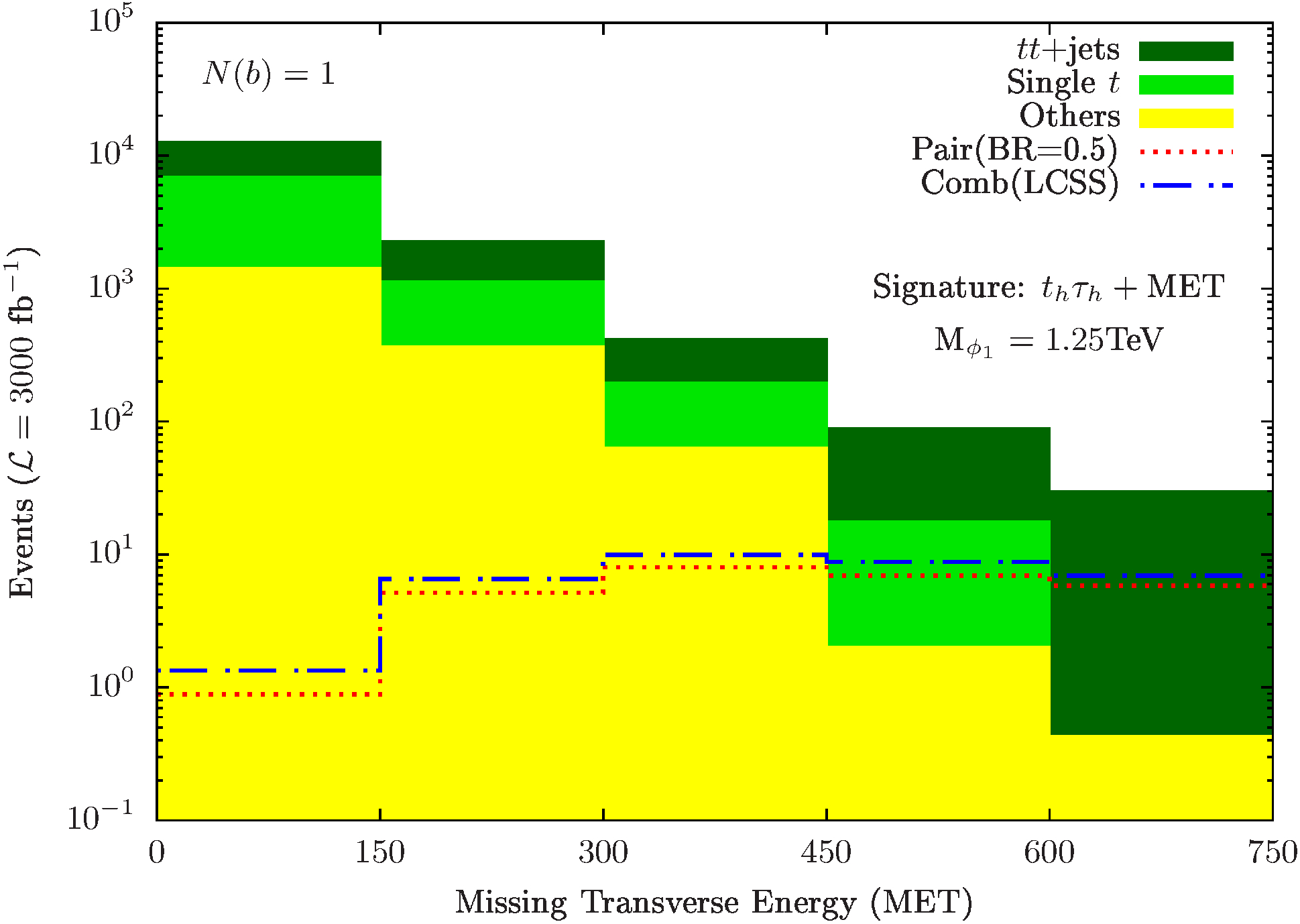}\label{fig:sig2_nbj1}}\quad\quad\quad\quad
\subfloat[\quad\quad\quad(d)]{\includegraphics[width=0.9\columnwidth, height=4.5cm]{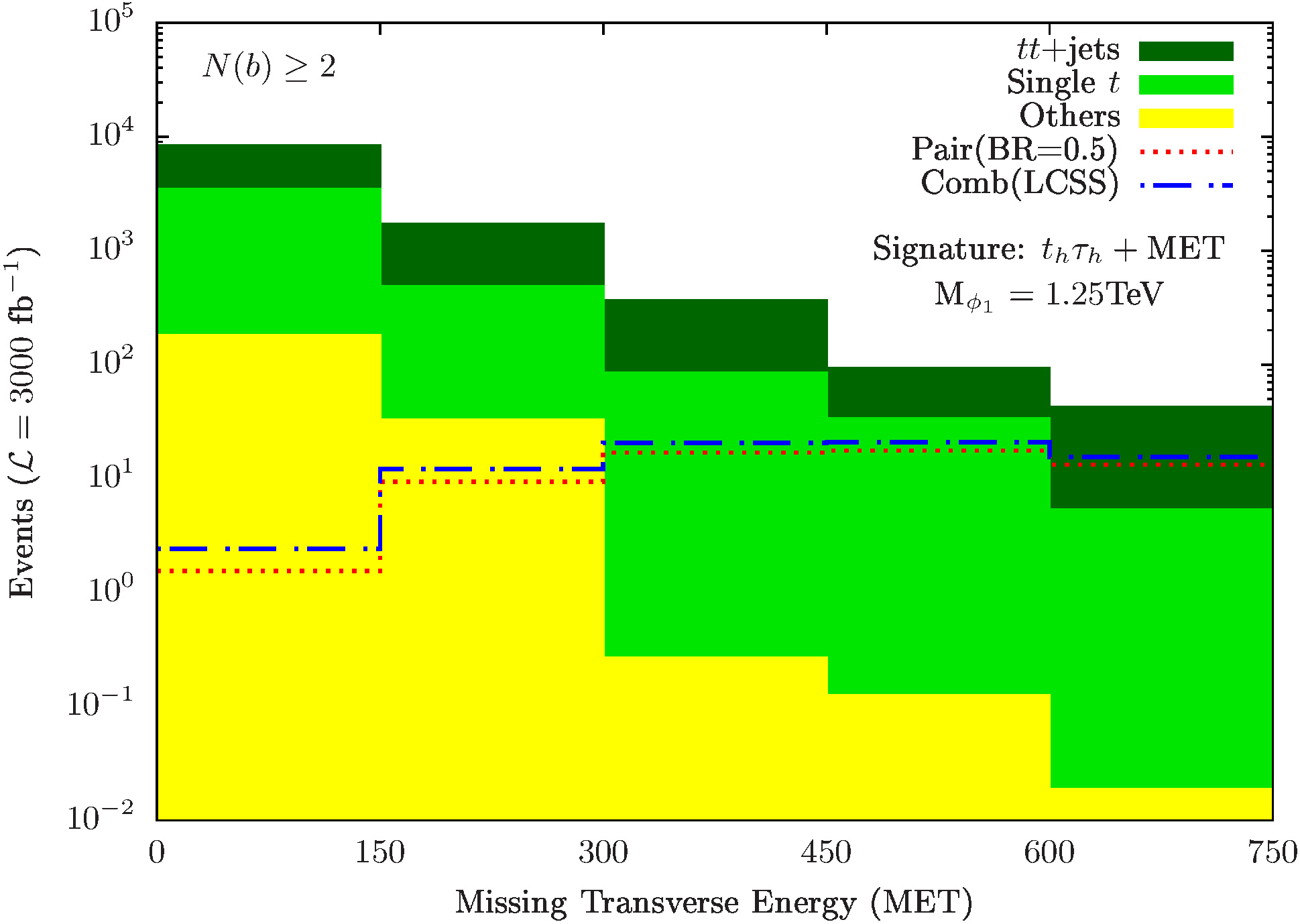}\label{fig:sig2_nbj2}}\\
\caption{The bins show the number of signal and background events in the $t_h\tau_h\tau_h + t_h\tau_h\tau_\ell$ channel [(a) and (b)] and the $t_h\tau_h+$ MET channel [(c) and (d)]. The events are obtained applying all the cuts in Table~\ref{tab:cuts}, except the ones on the variables used for binning, i.e., $N(\ta_h)$ and $M_{\rm T}$ in the  $t_h\tau_h\tau_h + t_h\tau_h\tau_\ell$ channel and $N(b)$ and MET in the $t_h\tau_h+$ MET channel.
}
\label{fig:binnedNbg_Nsig}
\end{figure*}

\begin{table*}[]
\centering{\linespread{3}
\begin{tabular*}{\textwidth}{c ||@{\extracolsep{\fill}} c c | c c | c c | c c ||  c c | c c }
\hline
 \multirow{2}{*}{~\rotatebox[origin=c]{90}{Comb. $\mc Z$ }} & \multicolumn{12}{c}{Limit on $M_{\phi_{1}}$ (TeV)}\\ 
	& \multicolumn{8}{c}{(Signature:~$t_h\tau_h\tau_h+t_h\tau_h\tau_\ell$)}	& \multicolumn{4}{c}{(Signature:~$t_h\tau_h+\textrm{MET}$)} \\ 
\cline{2-13}
&\multicolumn{2}{c|}{LCSS-Combined}&\multicolumn{2}{c|}{Pair (BR=$0.5$)}&\multicolumn{2}{c|}{RC-Combined}&\multicolumn{2}{c||}{Pair (BR=1)}&\multicolumn{2}{c|}{LCSS-Combined}&\multicolumn{2}{c}{Pair (BR=$0.5$)}\\
\cline{2-13}
& 5\% & 10\% & 5\% & 10\% & 5\% & 10\% & 5\% & 10\% & 5\% & 10\% & 5\% & 10\%\\
\hline\hline
~~5 & 1.06 & 0.94 & 0.84 & 0.75 & 1.42 & 1.37 & 1.40 & 1.36 & 1.19 & 1.11 & 1.15 & 1.05\\ \hline
~~3 & 1.34 & 1.26 & 1.12 & 1.04 & 1.56 & 1.51 & 1.54 & 1.49 & 1.37 & 1.30 & 1.32 & 1.24 \\ \hline
~~2 & 1.53 & 1.47 & 1.27 & 1.22 & 1.67 & 1.65 & 1.66 & 1.63 & 1.47 & 1.43 & 1.44 & 1.38 \\ \hline
\end{tabular*}}
\caption{The mass limits for some sample scenarios, with 5\% and 10\% systematic uncertainties.}
\label{tab:binnedsig1}
\end{table*}

\appendix
\section{Effect of systematic errors}\label{sec:appendixA}
\noindent 
Here, we show the effect of systematic uncertainty on our results. For illustration, we consider two benchmark choices of $5\%$ and $10\%$ systematics on our background estimations. In the presence of total systematic error $\sg_B$, Eq.~\eqref{eq:sig} generalises to
\begin{align}
\mathcal{Z}=&\sqrt2\Bigg((N_S+N_B)\ln\left[\frac{(N_S+N_B)(N_B+\sigma^2_B)}{N_B^2+(N_S+N_B)\sigma^2_B}\right]\nn\\
&\qquad-\left(\frac{N_B}{\sigma^2_B}\right)^2\ln\left[1+\frac{\sigma^2_B\, N_S}{N_B(N_B+\sigma^2_B)} \right]\Bigg)^{1/2},\label{eq:fullsignifacnce}
\end{align} 
whose approximated form is perhaps more familiar,
\begin{align}
\mathcal{Z}\approx\frac{N_S}{\sqrt{N_B+\sigma^2_B}}.
\end{align}
 A na\"ive use of the above formulae would adversely affect our estimation of the mass limits, especially in the cases of sLQs where the signal cross sections are smaller than the vLQs~\cite{Gross:2007zz}. However, so far, we have not utilised the signal and background distributions beyond the total number of events. Actually, it is possible to obtain comparable  mass limits even in the presence of $5\%$--$10\%$ systematic errors from binned data.

For an illustration, we show the limits estimated from a binned-data analysis for a few benchmark scenarios of $\ph_1$ in Table~\ref{tab:binnedsig1}. The numbers, obtained for both $5\%$ and $10\%$ systematic errors, are to be compared with those in Table~\ref{tab:sig2}. The bins are shown in Fig.~\ref{fig:binnedNbg_Nsig}.  For the vLQs, we find similar to slightly improved numbers ($\sim 5\%$--$10\%$) than those in Tables~\ref{tab:sig1} and~\ref{tab:sig3}.

To estimate the signal significance from binned data, we have employed the Liptak-Stouffer (weighted) $\mc Z$-score method where 
the metascore or the combined significance is given as
\begin{equation}
    \mathcal{Z}=\frac{\sum_{i=1}^{N}w_i \mc Z_i}{\sqrt{\sum_{i=1}^{N}w_i^2}}.
\end{equation} 
Here, $\mc Z_i$ denotes the signal significance in the $i^{\rm th}$ bin ($i\in\{1, 2, 3, \ldots, N\}$) computed from Eq.~\eqref{eq:fullsignifacnce} and $w_i$ is the corresponding weight, which is taken to be equal to the inverse of the variance in that bin. We have set the  $w_i$'s equal to the inverse of the square of the total errors, i.e., $w_i^{-1} = ({\rm statistical~error})^2+({\rm systematic ~error})^2=N_B^i+(\sg_B^i)^2$.

\bibliography{Leptoquark}{}
\bibliographystyle{JHEPCust}

\end{document}